\documentclass[
 aip,
 jcp,
 amsmath,amssymb,
 preprint
]{revtex4-2}

\usepackage{graphicx}
\usepackage{dcolumn}
\usepackage{booktabs}
\usepackage{bm}
\usepackage{amsmath}

\usepackage{placeins}
\usepackage[T1]{fontenc}
\usepackage[utf8]{inputenc}

\newcommand{\aipalt}[1]{\par\smallskip{\footnotesize\textbf{Alt text:} #1\par}\smallskip}

\begin{document}

\title{From Projected Subspaces to Full-Space Implementations:
Representation-Equivalence Audits for Open-Shell ADAPT-VQE}

\author{Lucia Mal\'{\i}\v{c}kov\'a}
\email{lucy@thesafehaven.ai}
\affiliation{Modelos Inteligencia Artificial S.L., Cl. Tajinaste 54, 386 20 San Miguel de Abona, Santa Cruz de Tenerife, Spain}

\author{Petr Klenovsk\'y}
\email{klenovsky@physics.muni.cz}
\affiliation{Department of Condensed Matter Physics, Faculty of Science, Masaryk University, Kotl\'a\v{r}sk\'a 267/2, 611 37 Brno, Czech Republic}
\affiliation{Czech Metrology Institute, Okru\v{z}n\'i 31, 638 00 Brno, Czech Republic}

\date{\today}

\begin{abstract}
A variational state optimized in a symmetry-projected space need not be the state prepared by exponentiating the corresponding unprojected parent generators.
For an orthogonal target-space projector $P$, $Q=I-P$, and an anti-Hermitian generator $A$, exact equivalence for all amplitudes requires target-space invariance, $QAP=0$.
We formulate this condition as a representation-equivalence audit for open-shell ADAPT-VQE.
For a canonical CAS(15e,9o) type-1 Cu(II) Hamiltonian, 274 of 323 bare Hartree--Fock-origin single, double, and triple excitations violate this condition.
Projected-doublet ADAPT reaches 1.5719 m$E_h$ with 121 operators, whereas the same amplitudes in the corresponding full-space sequence give 464.730 m$E_h$ error, 0.07343 fidelity with the projected state, and 54.45\% quartet weight.
Reoptimizing the same 121 parent generators in the same order under the hard constraint $p_D\ge0.999$ reduces the best converged error to 8.482 m$E_h$, showing that the catastrophic same-angle failure is primarily a representation/parameter-transfer failure but is not completely repaired by full-space reoptimization of the fixed sequence.
An independent 12-qubit NO benchmark exhibits the same mechanism more mildly.
A spin-preserving generalized full-space construction removes the representation ambiguity and reaches 1.493 and 1.181 m$E_h$ at 18 and 24 qubits.
For the 18-qubit positive control, selectively refining only four Suzuki-2 factors and re-transpiling with a fixed compiler seed passes the predeclared circuit-equivalence gate at 19,660 controlled-X (CX) gates and depth 23,016.
Circuit synthesis and final-energy measurement therefore constitute separate validation layers.
The contribution is a validation framework, not a new spin-adapted ADAPT algorithm.
\end{abstract}

\maketitle

\section{Introduction}

Variational quantum eigensolvers (VQEs) and ADAPT-VQE are widely studied approaches to quantum electronic-structure simulation \cite{peruzzo2014vqe,mcclean2016theory,mcardle2020review,grimsley2019adapt}.
Qubit pools, symmetry-aware and symmetry-complete pools, spin projection, compact excitation circuits, and pruning have all been developed subsequently \cite{tang2021qubitadapt,shkolnikov2023roadblocks,bertels2022symmetry,tsuchimochi2022spinproj,magoulas2023cnot,vaquero2025pruned}.
For open-shell systems, fixed particle number and $M_S$ do not determine total spin, and symmetry preservation can be essential for state targeting \cite{gard2020symmetry,seki2020symmetry,seki2022projection,burton2023disco,magoulas2026symmetry}.
Likewise, finite product-formula implementations of spin-adapted generators can introduce symmetry error, motivating exact or closed-form factorizations \cite{magoulas2025closedform,jain2026factorization}.

Several neighboring approaches therefore solve important but distinct problems.
Symmetry projection restores or enforces a desired sector in the variational state or objective \cite{seki2020symmetry,tsuchimochi2022spinproj,seki2022projection}; symmetry-preserving state-preparation circuits and symmetry-aware or symmetry-complete pools constrain the accessible full-space manifold \cite{gard2020symmetry,shkolnikov2023roadblocks,burton2023disco,magoulas2026symmetry}; contextual-subspace VQE deliberately defines a reduced-space Hamiltonian and projected operator pool \cite{weaving2023stabilizer}; and recent exact or closed-form factorizations address how a specified spin-adapted full-space unitary is compiled \cite{magoulas2025closedform,jain2026factorization}.
None of these statements by itself establishes that, for an arbitrary parent generator $A$, the reduced-space unitary $\exp(\theta U^\dagger A U)$ is identical to the target-space compression of the full-space parent evolution $\exp(\theta A)$.
The present work isolates and audits precisely this representation-identification step.

\begin{figure*}[htbp]
\centering
\includegraphics[width=0.98\textwidth]{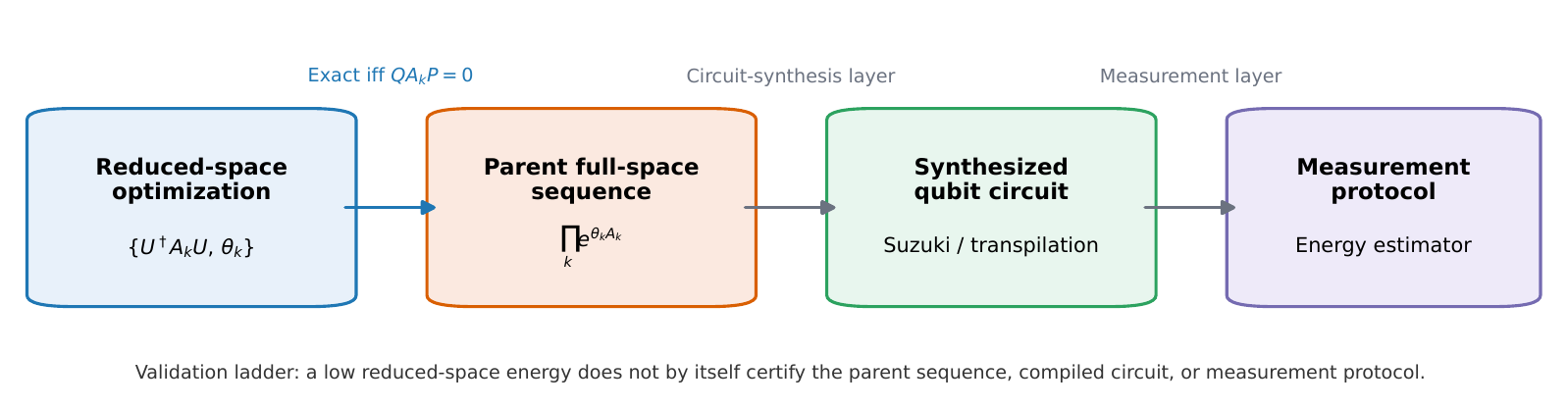}
\caption{Validation layers that should not be identified without an explicit check. The first arrow is exact for all amplitudes only when each implemented parent generator preserves the target subspace, $QA_kP=0$. Subsequent arrows add independent circuit-synthesis and measurement approximations.}
\label{fig:validation_ladder}
\end{figure*}

Projected adaptive pools are not new.
For example, contextual-subspace VQE optimizes projected operators directly in a reduced space \cite{weaving2023stabilizer}.
Such algorithms are legitimate reduced-space variational methods.
The narrower issue considered here arises only if the corresponding unprojected parent-generator sequence is subsequently treated as the full-space implementation, or assigned a circuit resource count, without proving that it realizes the same variational state.
Operationally, the ambiguity can enter when (i) parent excitations are projected or compressed into a target sector, (ii) ADAPT selection and parameter optimization are performed in that reduced representation, (iii) the selected parent labels and amplitudes are retained, and (iv) the unprojected parents are later exponentiated to construct a fermionic or qubit circuit or to quote resources.
The reduced calculation never explores the complementary $Q$ sector, so a low reduced-space energy cannot diagnose a nonzero $QAP$.
Figure~\ref{fig:validation_ladder} summarizes the resulting validation ladder.
The underlying invariant-subspace algebra is elementary; the practical question is how large the mismatch can become in an adaptive workflow and how to detect it before reduced-space optimization, full-space evolution, circuit synthesis, and measurement are conflated.

The practical significance is therefore deliberately narrower than a criticism of projected or symmetry-adapted methods themselves.
A reduced-space optimization certifies the variational object defined in that reduced representation.
Associating it with a particular parent full-space fermionic sequence, compiled circuit, or resource estimate is an additional statement that requires its own representation-equivalence check.
We do not assume that prior projected-space studies make this identification; rather, we isolate the condition that must be verified whenever such a connection is made.

The primary benchmark is an idealized type-1 (T1, ``blue-copper'') Cu(II) first-shell model, chosen because it provides a chemically motivated open-shell Hamiltonian while retaining exact spin-resolved references \cite{solomon2004bluecopper,solomon2014copper}.
A separate NO radical benchmark and Hubbard-model stress tests probe whether the same representation issue persists beyond that Hamiltonian.
No chemically converged prediction for blue-copper proteins, NO, or OH is claimed.

\section{Model and Validation Strategy}

The canonical Cu Hamiltonian is defined in a complete active space, CAS(15e,9o), comprising 15 active electrons in 9 active spatial orbitals and mapped to 18 qubits.
Its STO-3G orbitals were generated from restricted open-shell Hartree--Fock (ROHF), followed by atomic-valence active-space (AVAS) selection targeting the Cu $3d$ and short Cu--S donor $3p$ manifolds and a spin-constrained complete-active-space self-consistent-field (CASSCF) calculation; the exact geometry and numerical settings are given in the supplementary material \cite{sun2018pyscf,sayfutyarova2017avas}.
An archived CAS(21e,12o) FCIDUMP is used only as a fixed 24-qubit algorithmic benchmark because its dedicated 12-orbital selection driver was not preserved.
The independent NO test uses ROHF/STO-3G CAS(7e,6o) at $R_{\rm NO}=1.1508$~\AA{}.

For the target doublet in the fixed-$M_S=1/2$ sector,
\begin{equation}
\mathcal H_D=\ker\!\left(\hat S_+\big|_{\mathcal H_{M_S=1/2}}\right),
\qquad
P=UU^\dagger,\quad Q=I-P,
\end{equation}
where the columns of $U$ form an orthonormal basis of $\mathcal H_D$.
For a full-space anti-Hermitian fermionic generator $A$, let
\begin{equation}
A_D=U^\dagger A U.
\end{equation}
The lifted reduced-space unitary and the compressed full-space evolution satisfy
\begin{equation}
Ue^{\theta A_D}U^\dagger
=
Pe^{\theta A}P
\quad\text{for all real }\theta
\end{equation}
if and only if
\begin{equation}
QAP=0.
\label{eq:invariance}
\end{equation}
For anti-Hermitian $A$, Eq.~\ref{eq:invariance} is equivalent to $[A,P]=0$.
If invariance fails,
\begin{align}
Qe^{\theta A}P&=\theta QAP+\mathcal O(\theta^2),\\
Pe^{\theta A}P-e^{\theta PAP}P
&=
-\frac{\theta^2}{2}(QAP)^\dagger(QAP)+\mathcal O(\theta^3).
\end{align}
The supplementary material gives the proof, finite-amplitude bounds, and a state-specific telescoping certificate.
The exact projector is used only as a classical validation oracle for these small benchmarks. For the $M_S$-preserving generators considered here, $[A,\hat S^2]=0$ is a projector-free sufficient condition for preserving the target total-spin sector.

The corrected ansatz uses full-space anti-Hermitian generators $B_k$ whose target-space leakage and normalized $S^2$ commutator defect vanish to numerical precision,
\begin{equation}
|\Psi(\boldsymbol\theta)\rangle
=
e^{\theta_M B_M}\cdots e^{\theta_1B_1}|\Phi_{\rm HF}\rangle.
\end{equation}
Pool construction, ADAPT selection, repeated operator use, optimization, pruning, circuit synthesis, and measurement allocation are documented in the SI.
Throughout, 1.6 m$E_h$ is only an internal benchmark for a specified active-space Hamiltonian, not a claim of chemical accuracy for the underlying molecular model.

\section{Results and Discussion}

\subsection{Cu: representation failure and fixed-sequence reoptimization}

The regenerated 121-operator projected-doublet ADAPT sequence reaches $E=-2518.987495464~E_h$, 1.5719 m$E_h$ above the exact target doublet.
Using the same selected parent generators and the same amplitudes in the full fixed-$M_S$ space instead gives 464.730 m$E_h$ error, $\langle S^2\rangle=2.383537$, doublet weight 0.45549, and only 0.07343 fidelity with the projected state.
Post-projecting that full-space state does not repair the result: its normalized doublet component remains 465.164 m$E_h$ above the target and has fidelity 0.16121 with the projected state.

This mismatch is not caused by one exceptional operator.
Of 323 bare Hartree--Fock-origin single, double, and triple excitations, 274 violate Eq.~\ref{eq:invariance}; the selected sequence contains 114 non-invariant factors.
The supplementary material sequence certificates are correspondingly non-informative for this severe case.

To distinguish a failure of parameter transfer from a failure of the fixed parent sequence itself, we next reoptimized all 121 amplitudes in the \emph{same generator order} directly in the full fixed-$M_S$ space while enforcing
\begin{equation}
p_D=\|U^\dagger\Psi\|^2\ge0.999.
\label{eq:hardspin}
\end{equation}
Starting from zero amplitudes and three independent Gaussian perturbations of that start, all constrained Sequential Least Squares Programming (SLSQP) runs converged; the best three solutions are indistinguishable at 8.4815 m$E_h$, and the four-run range is 8.4815--8.9734 m$E_h$.
The best state has $p_D=0.999$, $\langle S^2\rangle=0.753$, and fidelity 0.97989 with the original projected state.
Its normalized doublet component is still 7.335 m$E_h$ above the exact doublet.
Tightening the constraint to $p_D\ge0.9999$ gives 11.270 m$E_h$ from the zero-amplitude start.
Separate starts near the original projected amplitudes converge to poorer local minima, demonstrating nonconvexity; no global-optimality claim is made.

We also optimized the same fixed sequence directly for overlap with the projected state.
Starting from the best energy-constrained basin, unconstrained fidelity maximization converges reproducibly to
$F_{\rm proj}=0.988268$ with $p_D=0.99716$.
Imposing $p_D\ge0.999$ during the fidelity maximization gives
$F_{\rm proj}=0.987416$ from three tested starts.
These are converged local optima rather than certified global maxima, but none of the tested full-space searches reproduces the projected state exactly.
The principal same-sequence comparisons are summarized in Table~\ref{tab:parent_reopt}.

\begin{table}[htbp]
\centering
\small
\caption{Canonical Cu representation test for the same 121 parent generators. The constrained rows reoptimize all amplitudes in the full fixed-$M_S$ space without changing operator order.}
\label{tab:parent_reopt}
\begin{tabular}{lrrrr}
\toprule
Construction & $\Delta E$ (m$E_h$) & $p_D$ & $F_{\rm proj}$ & Post-proj.\ $\Delta E_D$ (m$E_h$)\\
\midrule
Projected reduced-space ansatz & 1.5719 & 1 & 1 & 1.5719\\
Same amplitudes, full space & 464.730 & 0.45549 & 0.07343 & 465.164\\
Full-space energy reopt., $p_D\ge0.999$ & 8.4815 & 0.99900 & 0.97989 & 7.3355\\
Full-space fidelity reopt., $p_D\ge0.999$ & 10.8166 & 0.99900 & 0.98742 & 9.8863\\
Full-space energy reopt., $p_D\ge0.9999$ & 11.2696 & 0.99990 & 0.97017 & 11.0305\\
\bottomrule
\end{tabular}
\end{table}

Thus the catastrophic 464.7 m$E_h$ failure is primarily a representation/parameter-transfer failure: the same parent sequence can recover much of the target state after full-space spin-constrained reoptimization.
However, in the tested order it still does not reproduce the 1.6 m$E_h$ projected result.
This distinction is central: the audit does not show that projected-pool methods are invalid, nor that the parent sequence is globally incapable of representing the target; it shows that the projected optimization and the unprojected same-parameter implementation are different variational constructions.

\begin{figure}[htbp]
\centering
\includegraphics[width=0.92\linewidth]{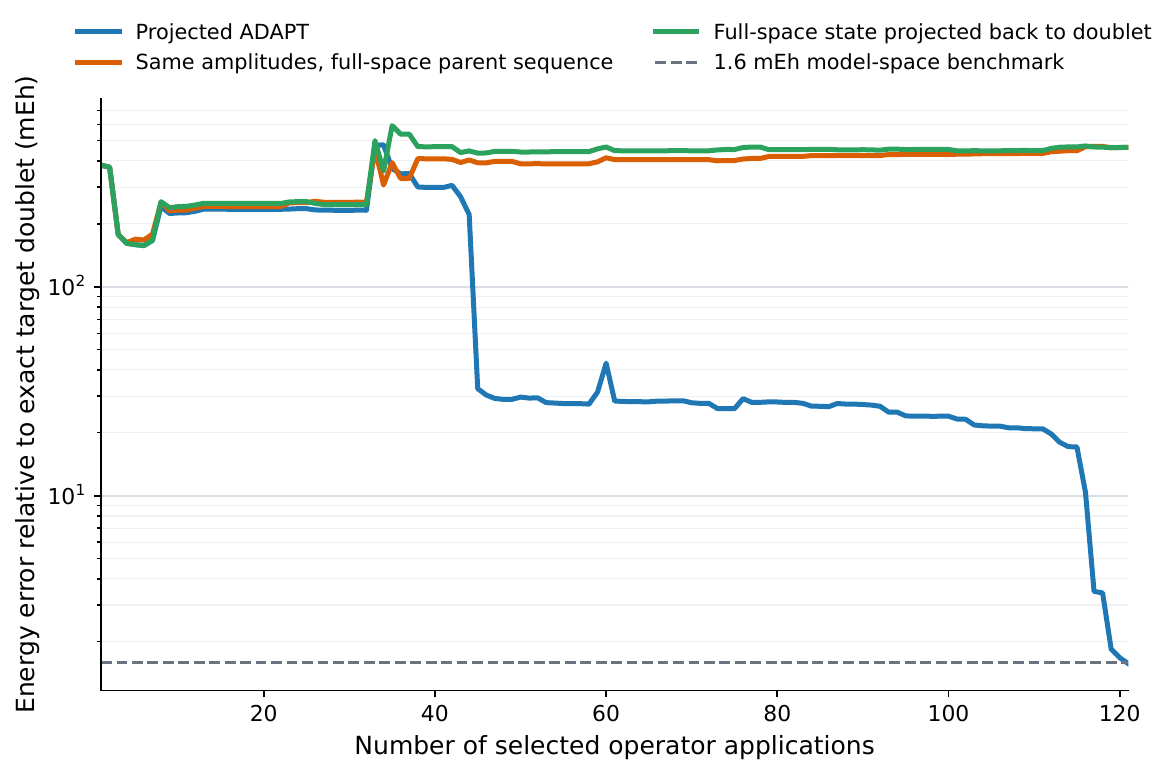}
\caption{Canonical Cu negative control. The projected energy approaches the target while the same amplitudes in the parent full-space sequence do not. Post-projection of the resulting full-space state does not recover the optimized projected trajectory.}
\label{fig:prefix}
\end{figure}

Simple invariant filtering is not an adequate repair.
Only 49 of the 323 Cu excitations are individually invariant.
Using all 49 once gives 210.619 m$E_h$ error, and allowing repeated selection through 100 applications gives 167.340 m$E_h$.
At the Hartree--Fock reference, the projected bare S/D/T pool spans all 239 real tangent directions of the 240-dimensional doublet space, whereas the invariant-only subset spans 49 and the generalized spin-preserving full-space rank-1/rank-2 pool spans 99.
These are local first-order diagnostics, not controllability statements.
The divergence of the projected, full-space, and post-projected trajectories over all 121 prefixes is shown in Fig.~\ref{fig:prefix}.

\subsection{Independent NO test}

For NO CAS(7e,6o), the exact target doublet lies 250.051 m$E_h$ below the lowest quartet, so representation failure cannot be attributed to collapse into a lower wrong-spin state.
Nevertheless, 225 of 231 bare S/D/T excitations are non-invariant.
Projected ADAPT reaches 1.4286 m$E_h$ with 16 operators; the same parent sequence and amplitudes give 24.537 m$E_h$, doublet weight 0.97211, and fidelity 0.97205 with the projected state.
Here the state-specific SI certificate is nontrivial, guaranteeing $F\ge0.8473$.
Unlike Cu, post-projection almost repairs the result: the post-projected error is 1.4780 m$E_h$ and its fidelity with the projected state is 0.999943.
The same mechanism therefore occurs in an independent molecular Hamiltonian, but its severity is strongly system and sequence dependent.

Only six NO parent excitations are individually invariant and this filtered pool remains 52.154 m$E_h$ above the target.
A generalized spin-preserving pool instead gives a 19-application state at 1.5790 m$E_h$, unit doublet weight, and 0.998967 overlap with the exact lowest-doublet manifold.
Additional OH and Hubbard stress tests are reported in the supplementary material.

\subsection{Representation-faithful full-space ans\"atze}

The generalized spin-preserving pool used below is a representation-faithful \emph{positive control}, not a claim of a new or preferred scalable spin-adapted ADAPT algorithm.
Existing work already provides symmetry-preserving state-preparation circuits, spin-adapted pools, symmetry-completeness analyses, and exact spin-adapted factorizations \cite{gard2020symmetry,shkolnikov2023roadblocks,burton2023disco,magoulas2026symmetry,magoulas2025closedform,jain2026factorization}.
Its purpose here is narrower: to demonstrate on the same benchmarks that, once every implemented generator preserves the target sector, the full-space and reduced-space evolutions do represent the same variational construction to numerical precision.

For Cu, the generalized full-space spin-preserving pools contain 1,080 generators at 18 qubits and 3,762 at 24 qubits.
After repeated selection, reoptimization, and pruning, the stored 18q and 24q ans\"atze contain 50 and 61 applications and reach 1.493 and 1.181 m$E_h$, respectively.
Independent full-space and reduced-space reconstructions agree to machine precision over every prefix.
The final representation-faithful Cu ans\"atze are summarized in Table~\ref{tab:main}.

\begin{table}[htbp]
\centering
\small
\caption{Compact representation-faithful Cu ans\"atze. The 24q case is a fixed-Hamiltonian algorithmic benchmark, not a chemical convergence study.}
\label{tab:main}
\begin{tabular}{lcc}
\toprule
& 18 qubits & 24 qubits\\
\midrule
Active space & CAS(15e,9o) & CAS(21e,12o)\\
Operator applications & 50 & 61\\
Unique generators & 39 & 41\\
$\Delta E$ (m$E_h$) & 1.4927 & 1.1813\\
Exact-doublet fidelity & 0.9826 & 0.9907\\
Residual ($E_h$) & 0.0480 & 0.0585\\
$\langle S^2\rangle$ & 0.7500 & 0.7500\\
\bottomrule
\end{tabular}
\end{table}

The spin-preserving pool basis is not mathematically unique inside multidimensional null spaces.
A normalization-controlled supplementary-material audit finds modest top-group sensitivity along the stored 18q trajectory: random normalized basis changes affect the highest-gradient group at only 4 of 51 sampled prefixes, while a basis-invariant group-gradient score agrees with the canonical choice at 49--50 of 51 prefixes depending on normalization metric.
No basis-independent minimality claim is made for the compact sequences.

\subsection{Circuit and measurement validation}

Second-order Suzuki synthesis introduces a separate approximation.
The 0.1 m$E_h$ synthesis-error threshold was fixed before the selective-refinement search so that circuit-synthesis error would contribute at most 6.25\% of the 1.6 m$E_h$ model-space accuracy benchmark, leaving the ansatz error as the dominant contribution.
For the 18q Cu state, four uniform repetitions give 0.135 m$E_h$ synthesis error and circuit-to-fermionic fidelity 0.999484 at 61,482 CX gates and abstract-gate-set depth 71,636, so this uniform circuit narrowly fails the predeclared $0.1$ m$E_h$ synthesis-error gate.
The factorization error is highly nonuniform, however.
A restricted emulator search identifies the schedule $r_9=4$, $r_{21}=r_{25}=r_{34}=2$, with all other applications at $r=1$.
Re-transpiling exactly this schedule with Qiskit 2.2.3, optimization level 1, the abstract basis $\{R_z,\sqrt X,X,\mathrm{CX}\}$, and fixed \texttt{seed\_transpiler}=9272026 gives
$|\delta E_{\rm synth}|=0.095021$ m$E_h$,
$F_{\rm circuit,fermionic}=0.9995825$,
$\langle S^2\rangle=0.750724$,
and target-$M_S$ weight $0.9999999999994$.
It therefore passes all four predeclared state-level gates at 19,660 CX gates and depth 23,016, reductions of about 68.0\% and 67.9\%, respectively, relative to uniform $r=4$.
For 24q, one repetition gives 0.0906 m$E_h$ and fidelity 0.999694 at 22,574 CX gates and depth 26,409; for NO, one repetition gives 0.00293 m$E_h$ and fidelity 0.9999939.
All reported resources are abstract-gate-set compilations without a device coupling map and are not optimal-circuit or hardware-resource bounds.
Further schedule and provenance details are given in the supplementary material.

Final-energy measurement is likewise protocol dependent.
For the chosen greedy qubit-wise commuting (QWC) groupings and ideal state preparation, variance-optimal allocation requires approximately $1.43\times10^8$ shots (18q) and $5.46\times10^8$ shots (24q) for a 1.6 m$E_h$ statistical uncertainty.
These numbers exclude ADAPT gradient measurements, optimizer evaluations, device noise, and alternative measurement strategies.

\subsection{Scope of claims}

Spin projection, symmetry-preserving state preparation, projected pools, generalized excitations, repeated selection, spin-adapted or symmetry-complete pools, and pruning are established ideas \cite{gard2020symmetry,seki2020symmetry,tsuchimochi2022spinproj,seki2022projection,weaving2023stabilizer,shkolnikov2023roadblocks,burton2023disco,magoulas2026symmetry}.
Our contribution is not a new projector, a new general spin-adapted pool, or a new factorization method, but a representation-equivalence validation framework for connecting a reduced-space optimization to a parent full-space implementation or resource estimate.
The exact projector used here serves only as a small-system validation oracle.

The molecular calculations are fixed algorithmic benchmarks rather than chemically converged predictions.
The compact ADAPT sequences are search-history dependent, and the reported circuit and measurement resources are specific to the documented synthesis, transpilation, and measurement protocols; no global minimality, hardware-optimality, or quantum-advantage claim is made.

\section{Conclusions}

A projected adaptive ansatz and the corresponding parent full-space sequence are guaranteed to define the same generator-by-generator variational construction for arbitrary amplitudes when every implemented parent generator preserves the target space.
The converse necessity applies to this amplitude-independent representation guarantee; accidental agreement at isolated parameter values or cancellations within one particular finite sequence are not excluded.
The canonical Cu example shows how severely this identification can fail: 1.5719 m$E_h$ in projected space becomes 464.730 m$E_h$ when the same amplitudes are applied to the parent sequence.
Full-space energy reoptimization of that fixed sequence under $p_D\ge0.999$ reduces the best converged error to 8.482 m$E_h$ and raises the projected-state fidelity to 0.97989.
Direct fidelity maximization under the same purity constraint reaches 0.98742 but still does not reproduce the projected state.
The catastrophic failure is therefore not evidence that the parent sequence is intrinsically useless; it is primarily evidence that reduced-space parameters cannot be transferred without an equivalence check.
The remaining energy and fidelity gaps show that the fixed-order full-space construction does not completely recover the projected construction in the tested searches, without establishing a global impossibility result.

An independent NO benchmark reproduces the mechanism more mildly, while representation-faithful spin-preserving Cu and NO ans\"atze provide positive controls.
For the 18q Cu positive control, a selectively refined Suzuki-2 schedule also passes the predeclared circuit-equivalence gate after an actual Qiskit re-transpilation, with 19,660 CX gates and depth 23,016 in the abstract basis.
Circuit synthesis and measurement therefore remain separate approximations and resource layers rather than consequences of fermionic-level accuracy.

The practical recommendation is correspondingly narrow:
when a projected or otherwise reduced-space adaptive optimization is connected to a full-space implementation, the optimized state, the parent fermionic sequence, the compiled circuit, and the measurement protocol should be validated as distinct objects.
A low reduced-space energy alone does not establish that a quoted full-space circuit implements the same variational state.

\section*{SUPPLEMENTARY MATERIAL}
See the supplementary material for the complete Hamiltonian-generation protocol; proofs and finite-amplitude bounds; numerical tolerances; state-specific certificates; fixed-parent energy and fidelity reoptimization; Cu prefix and pool audits; tangent-space accessibility; normalization-controlled basis-sensitivity tests; NO, OH, and Hubbard benchmarks; spin-preserving pool construction; circuit and Suzuki diagnostics; measurement grouping; one-particle diagnostics; optimizer robustness; provenance; and excluded calculations.

\begin{acknowledgments}
L.M. gratefully acknowledges the use of the quantum system Euro-Q-Exa, co-funded by the EuroHPC JU, BMFTR, and the Bavarian State Ministry of Science and the Arts, operated by the Leibniz Supercomputing Centre (LRZ) in Garching, Germany, for providing the computational resources for this work. Access to the Euro-Q-Exa infrastructure was awarded under the EuroHPC project proposal No.~EHPC-QCP-2026Q01-017.

P.K. acknowledges the Institutional Subsidy for the Long-Term Conceptual Development of a Research Organization granted to the Czech Metrology Institute by the Ministry of Industry and Trade of the Czech Republic; and the project \emph{Quantum Materials for Applications in Sustainable Technologies}, CZ.02.01.01/00/22\_008/0004572.
\end{acknowledgments}

\section*{AUTHOR DECLARATIONS}

\subsection*{Conflict of Interest}
The authors have no conflicts to disclose.

\subsection*{Author Contributions}
Lucia Mal\'{\i}\v{c}kov\'a: Conceptualization (lead; selection of the research topic); Investigation (lead; execution of the quantum-computing calculations); Writing -- original draft (supporting).

Petr Klenovsk\'y: Formal analysis (lead; many-body calculations outside the quantum-computing workflow); Validation (lead; physical consistency checks); Investigation (supporting); Writing -- original draft (lead); Writing -- review \& editing (lead).

\section*{DATA AVAILABILITY}
The data that support the findings of this study are openly available in the public GitHub repository \url{https://github.com/lucia-malickova/New-VQE-validation}. The repository contains the Hamiltonians, checkpoints, representation-equivalence audits, fixed-parent energy- and fidelity-reoptimization tests, independent benchmarks, circuit outputs, and figure source data used in this work.

\bibliography{references}

\title{Supplementary Material for:
From Projected Subspaces to Full-Space Implementations:
Representation-Equivalence Audits for Open-Shell ADAPT-VQE}

\author{Lucia Mal\'{\i}\v{c}kov\'a}
\affiliation{Modelos Inteligencia Artificial S.L., Cl. Tajinaste 54, 386 20 San Miguel de Abona, Santa Cruz de Tenerife, Spain}

\author{Petr Klenovsk\'y}
\affiliation{Department of Condensed Matter Physics, Faculty of Science, Masaryk University, Kotl\'a\v{r}sk\'a 267/2, 611 37 Brno, Czech Republic}
\affiliation{Czech Metrology Institute, Okru\v{z}n\'i 31, 638 00 Brno, Czech Republic}

\date{\today}
\maketitle

\section{Benchmark Hamiltonians}

\subsection{Type-1 Cu active spaces}

The primary benchmark is an idealized oxidized first-coordination-shell type-1 Cu(II) model with two nitrogen donors, a cysteine-model sulfur donor, and a longer axial sulfur donor.
The molecular orbitals were obtained from a restricted open-shell Hartree--Fock (ROHF) calculation in the STO-3G basis.
The primary complete active space, CAS(15e,9o), contains 15 active electrons in 9 active spatial orbitals and was constructed with PySCF \cite{sun2018pyscf} and the atomic-valence active-space (AVAS) procedure \cite{sayfutyarova2017avas}, targeting the Cu $3d$ and cysteine-S $3p$ manifolds.
CAS(21e,12o) from the same model is used as an expanded algorithmic benchmark.

The active-space Hamiltonian is written
\begin{equation}
\hat H
=
E_{\rm core}
+\sum_{pq} h_{pq} a_p^\dagger a_q
+\frac{1}{2}\sum_{pqrs}(pq|rs)
a_p^\dagger a_r^\dagger a_s a_q .
\label{eq:H}
\end{equation}
The exact Cu target-spin energies used throughout are
\begin{align}
E_D^{18}&=-2518.989067369946~E_h,\\
E_Q^{18}&=-2518.995009103114~E_h,\\
E_D^{24}&=-2518.989131438323~E_h,\\
E_Q^{24}&=-2518.995161621996~E_h.
\end{align}
The lower quartet is a property of these deliberately small approximate active-space Hamiltonians and is not interpreted as a prediction for the experimental oxidized T1 site.

\subsubsection{Canonical Cu geometry and Hamiltonian-generation protocol}

The canonical 18q FCIDUMP can be traced back to the archived geometry-to-Hamiltonian driver.
The idealized first-shell geometry (in \AA) is
\begin{center}
\begin{tabular}{crrr}
\toprule
Atom & $x$ & $y$ & $z$\\
\midrule
Cu &  0.000 &  0.000 &  0.000\\
N  &  2.000 &  0.000 &  0.000\\
H  &  2.340 &  0.900 &  0.200\\
H  &  2.340 & -0.620 &  0.740\\
H  &  2.340 & -0.280 & -0.800\\
N  & -1.000 &  1.732 &  0.000\\
H  & -1.280 &  2.200 &  0.800\\
H  & -1.550 &  2.120 & -0.620\\
H  & -0.120 &  2.260 &  0.040\\
S  & -0.600 & -1.100 &  1.750\\
H  & -1.420 & -1.580 &  2.120\\
S  &  0.200 & -0.500 & -2.830\\
H  &  0.920 & -1.180 & -3.200\\
H  & -0.540 & -0.920 & -3.280\\
\bottomrule
\end{tabular}
\end{center}
The model has total charge $+1$ and spin input $2S=1$.
The short Cu--S donor is represented by SH$^-$ and the longer axial donor by SH$_2$; the two nitrogen donors are NH$_3$ fragments.
These fragments are intentionally minimal and are used only to generate a controlled open-shell benchmark.

The archived PySCF driver uses STO-3G, ROHF with level shift 0.2 and at most 300 SCF cycles (with a second-order Newton fallback if needed), followed by AVAS with labels
\texttt{["Cu 3d","9 S 3p"]} and threshold 0.2.
In the zero-based atom indexing of the geometry above, atom 9 is the short Cu--S donor.
For the returned CAS(15e,9o), a spin-constrained CASSCF is then run with
\texttt{max\_cycle\_macro=50}, \texttt{max\_cycle\_micro=20},
energy tolerance $10^{-6}$, gradient tolerance $10^{-4}$,
FCI-solver tolerance $10^{-7}$, and \texttt{fix\_spin\_(ss=0.75)}.
The reduced one- and two-electron integrals are exported directly to FCIDUMP.
The generator script explicitly allows use of the best orbitals if the macro-iteration cap is reached; a separate convergence flag was not serialized with the archived FCIDUMP, so we do not claim a fully converged chemical CASSCF result from this model.
The scientific use of the file is instead as a fixed, exactly diagonalizable active-space Hamiltonian.

The expanded CAS(21e,12o) FCIDUMP is retained from the same model campaign and is exactly validated as a fixed 24-qubit Hamiltonian.
However, the separate 12-orbital selection driver that originally generated this expanded file is not preserved in the repository.
Accordingly, the 24q case is used only as an algorithmic benchmark and is not claimed to be reproducible from the geometry alone.

\subsection{Independent NO radical benchmark}

As an independent molecular test, an NO radical at $R_{\rm NO}=1.1508$~\AA{} was treated at the ROHF/STO-3G level.
Freezing the lowest four canonical spatial orbitals gives CAS(7e,6o), i.e.\ 12 spin orbitals.
The fixed-$M_S=1/2$ determinant space has dimension 300 and the exact doublet subspace dimension is 210.
The lowest exact doublet energy is
\begin{equation}
E_D^{\rm NO}=-127.632713244383~E_h,
\end{equation}
whereas the lowest quartet is higher by 250.051 m$E_h$.
The lowest doublet is a numerically twofold-degenerate manifold; the next doublet lies 296.438 m$E_h$ higher.
Accordingly, ground-manifold weights rather than fidelity to one arbitrary eigenvector are used when appropriate.

\subsection{Auxiliary OH radical stress test}

An OH radical with O at $(0,0,0)$~\AA{} and H at $(0,0,0.970)$~\AA{} was treated at ROHF/STO-3G, freezing the lowest spatial orbital to obtain CAS(7e,5o), i.e.\ 10 qubits.
The fixed-$M_S=1/2$ space has dimension 50 and the exact doublet subspace dimension is 40.
The lowest doublet is again numerically twofold degenerate; the next doublet lies 223.614 m$E_h$ higher.
This system is used only as an auxiliary stress test because the generalized physical ADAPT search discussed below does not reach the 1.6 m$E_h$ benchmark.

\subsection{Homogeneous Hubbard-chain stress tests}

To remove molecule-specific ingredients, we also use a homogeneous open Hubbard chain
\begin{equation}
\hat H
=
-t\sum_{\langle ij\rangle,\sigma}
\left(
a_{i\sigma}^\dagger a_{j\sigma}
+a_{j\sigma}^\dagger a_{i\sigma}
\right)
+
U\sum_i n_{i\uparrow}n_{i\downarrow}.
\end{equation}
The reported stress tests use four sites, five electrons, $M_S=1/2$, $t=1$, and
$U/t=1,2,4,8$.
They are algebraic validation models, not chemical benchmarks.

\section{Exact spin resolution}

Particle number and the spin projection $M_S$ do not by themselves determine the total spin $S$.
We therefore resolve the target doublet sector explicitly.
The spin-raising operator is
\begin{equation}
\hat S_+=\sum_p a_{p\alpha}^\dagger a_{p\beta},
\end{equation}
where $a_{p\sigma}^\dagger$ and $a_{p\sigma}$ create and annihilate an electron, respectively, in spatial orbital $p$ with spin $\sigma$.
Acting with $\hat S_+$ converts a $\beta$ electron into an $\alpha$ electron and raises $M_S$ by one.

Within the fixed-$M_S=1/2$ determinant space, a pure doublet is a highest-weight state and is therefore annihilated by $\hat S_+$.
The target doublet subspace is thus
\begin{equation}
\mathcal H_D
=
\ker\!\left(\hat S_+\big|_{\mathcal H_{M_S=1/2}}\right),
\end{equation}
where $\ker$ denotes the null space.
Let the columns of $U$ form an orthonormal basis of $\mathcal H_D$.
Then $U^\dagger U=I$ in the reduced doublet coordinates, and
\begin{equation}
P=UU^\dagger,\qquad Q=I-P
\end{equation}
are complementary orthogonal projectors in the full fixed-$M_S$ determinant space.
Specifically,
\[
P^2=P=P^\dagger,\qquad
Q^2=Q=Q^\dagger,\qquad
PQ=QP=0.
\]
The notation $\operatorname{Ran}(P)$ means the range (image) of $P$, i.e. the set of states onto which $P$ projects.
Here
\[
\operatorname{Ran}(P)=\mathcal H_D,
\qquad
\operatorname{Ran}(Q)=\mathcal H_D^\perp.
\]
Consequently, a state belongs entirely to the target doublet sector exactly when $P|\psi\rangle=|\psi\rangle$, or equivalently $Q|\psi\rangle=0$.
For a normalized state, $\|Q|\psi\rangle\|^2$ is therefore its weight outside the target subspace.

If $H_{M_S}$ denotes the Hamiltonian matrix in the full fixed-$M_S$ determinant basis, its exact representation in the doublet basis is
\begin{equation}
H_D=U^\dagger H_{M_S}U .
\end{equation}
For a normalized state in the $M_S=1/2$ sector,
\begin{equation}
\langle S^2\rangle
=
\|\hat S_+|\psi\rangle\|^2+M_S(M_S+1).
\end{equation}
Thus $\hat S_+$ provides both the exact doublet basis and a direct diagnostic of spin contamination.

The explicit matrix $U$ and projector $P=UU^\dagger$ are classical-oracle objects used because the present benchmark sectors are small enough to resolve exactly.
They are not proposed as scalable quantum-circuit primitives.
For spin, the production physical pool uses $M_S$-preserving excitation patterns and imposes the generator-level condition $[B,\hat S^2]\simeq0$; together these conditions preserve the target spin sector, while the exact-projector leakage is verified only as a classical benchmark.
For an arbitrary projected subspace lacking a compact symmetry generator, a scalable surrogate for $QAP=0$ is necessarily problem dependent.

\section{Numerical conventions, tolerances, and optimization settings}

Throughout the Supplement, $\|\cdot\|$ denotes the Euclidean norm for state vectors, $\|\cdot\|_2$ the spectral (operator) norm for matrices, and $\|\cdot\|_F$ the Frobenius norm.
For a full-space generator $A$, we quantify its normalized target-space leakage by
\begin{equation}
\ell_F(A)
=
\frac{\|QAU\|_F}{\|AU\|_F},
\end{equation}
whenever the target-space action in the denominator is nonzero.
The numerator is the part of $A$ that carries target-space basis vectors into the complementary $Q$ sector.

The exact highest-weight basis is constructed with \texttt{scipy.linalg.null\_space} using relative cutoff $10^{-12}$; the stored implementation rejects an $S_+$ null-space defect above $10^{-9}$.
Bare determinant-excitation generators are required to satisfy an anti-Hermiticity defect below $10^{-10}$.
For the generalized physical Cu pools, coefficient null spaces use relative cutoff $10^{-11}$; candidates with Frobenius norm or target-space action below $10^{-12}$ are discarded, and the accepted target-space leakage ratio and normalized $S^2$ commutator defect must each be below $2\times10^{-9}$.
The latter defect is defined as
\begin{equation}
\eta_{S^2}(B)=\frac{\|[\hat S^2,B]\|_F}{\|B\|_F}.
\end{equation}
The reported final pools are many orders of magnitude below these acceptance thresholds.
For the representation audits, a bare generator is classified as numerically non-invariant when $\ell_F>10^{-10}$.

The regenerated canonical Cu projected-ADAPT negative control uses the Hartree--Fock-origin rank-1/rank-2/rank-3 pool without repeated selection.
At each iteration the unused operator with the largest absolute energy derivative is chosen; ties follow the deterministic first-index behavior of \texttt{numpy.argmax}.
All selected amplitudes are globally reoptimized after every addition with the limited-memory Broyden--Fletcher--Goldfarb--Shanno algorithm with box constraints (L-BFGS-B), using \texttt{gtol}$=10^{-8}$, \texttt{ftol}$=10^{-12}$, at most 1000 iterations, and at most 50 line-search steps.
The run terminates when the projected target-doublet error first reaches 1.6 m$E_h$, which occurs at 121 applications.

For the Cu generalized physical ADAPT searches, repeated selection is allowed.
The search implementation reoptimizes the most recent 12 amplitudes after every addition with L-BFGS-B (maximum 80 iterations, \texttt{ftol}$=10^{-12}$, \texttt{gtol}$=10^{-8}$, maximum 30 line-search steps) and performs a global reoptimization every 20 additions (maximum 160 iterations, \texttt{ftol}$=10^{-13}$, \texttt{gtol}$=10^{-8}$, maximum 40 line-search steps), followed by a final global optimization with at least 300 iterations allowed.
Magnitude pruning retains the requested number of applications with the largest $|\theta_k|$ and then globally reoptimizes with L-BFGS-B using \texttt{ftol}$=10^{-14}$, \texttt{gtol}$=10^{-9}$, at most 80 iterations, 50 line-search steps, and memory parameter \texttt{maxcor}$=30$.
These settings describe the stored Cu search and pruning utilities; the reported sequence lengths are therefore search-history dependent rather than proofs of minimality.

All Hartree--Fock tangent ranks reported below are computed by numerical matrix rank with absolute tolerance $10^{-10}$.
Because the FCIDUMP integrals and determinant-space excitation matrices used in these audits are real, the relevant normalized target-state manifold is treated as real: for target dimension $d_D$, the maximum nontrivial tangent rank at one normalized state is therefore $d_D-1$.

\section{Representation-equivalence criterion}

Let $A^\dagger=-A$ be a full-space anti-Hermitian parent generator and define its reduced representation
\[
A_D=U^\dagger A U.
\]
There are then two different evolutions that must not be identified without a check.
The reduced-space evolution is $e^{\theta A_D}$; lifted back to the full determinant space it is
$Ue^{\theta A_D}U^\dagger$.
The actual parent evolution is $e^{\theta A}$; when only its target-space component is retained, it becomes $Pe^{\theta A}P$.
The question is therefore whether these two operators agree on $\operatorname{Ran}(P)$.

For clarity, relative to the orthogonal decomposition
\[
\mathcal H_{M_S}
=
\operatorname{Ran}(P)\oplus\operatorname{Ran}(Q),
\]
the generator has the block form
\[
A=
\begin{pmatrix}
PAP & PAQ\\
QAP & QAQ
\end{pmatrix}.
\]
The off-diagonal block $QAP$ maps a target-space state into the complementary subspace and is therefore the generator-level leakage block.
Because $A$ is anti-Hermitian,
$PAQ=-(QAP)^\dagger$, so the two off-diagonal blocks vanish together.

The following conditions are equivalent:
\begin{align}
QAP&=0,\\
[A,P]&=0,\\
e^{\theta A}\operatorname{Ran}(P)&\subseteq\operatorname{Ran}(P)
\quad\text{for all real }\theta,\\
P e^{\theta A}P
&=
Ue^{\theta A_D}U^\dagger
=
e^{\theta PAP}P
\quad\text{for all real }\theta.
\end{align}
The third condition states geometrically that the full-space evolution never carries a target-space state out of $\operatorname{Ran}(P)$.
The fourth condition states operationally that the lifted reduced-space unitary and the compressed parent full-space evolution are the same operator on the target sector.

If $QAP=0$, anti-Hermiticity also gives $PAQ=0$, so $A=PAP+QAQ$ is block diagonal with respect to the $P/Q$ decomposition.
Conversely, equality of the second derivatives of the last relation at $\theta=0$ gives
\begin{equation}
PA^2P-(PAP)^2
=
PAQAP
=
-(QAP)^\dagger(QAP),
\end{equation}
which vanishes only if $QAP=0$.
This proves necessity for equality at arbitrary amplitudes.

This necessity statement concerns equality of the generator-level reduced and compressed full-space evolutions for arbitrary amplitudes.
For a particular finite product at one isolated parameter vector, accidental agreement or cancellation between non-invariant factors is not excluded; such a coincidence would not constitute an amplitude-independent representation guarantee.

The short-amplitude expansions make the hierarchy explicit:
\begin{align}
Qe^{\theta A}P
&=
\theta QAP+\mathcal O(\theta^2),\\
Pe^{\theta A}P-e^{\theta PAP}P
&=
-\frac{\theta^2}{2}(QAP)^\dagger(QAP)
+\mathcal O(\theta^3).
\end{align}
Thus leakage \emph{amplitude} out of the target space starts at first order in $\theta$, whereas the mismatch between the two target-space evolutions starts at second order.

\subsection{Finite-amplitude operator-norm bounds}

Separate $A$ into its block-diagonal and block-off-diagonal parts,
\begin{equation}
A_0=PAP+QAQ,\qquad
A_{\rm off}=PAQ+QAP,
\end{equation}
so that $A=A_0+A_{\rm off}$.
The operator $A_0$ preserves the $P$ and $Q$ sectors separately, whereas $A_{\rm off}$ couples them.
For anti-Hermitian $A$,
\begin{equation}
\|A_{\rm off}\|_2=\|QAP\|_2\equiv\lambda.
\end{equation}
The interaction-picture Dyson series then gives the finite-amplitude bounds
\begin{align}
\|Qe^{\theta A}P\|_2
&\le\sinh(|\theta|\lambda),\\
\|Pe^{\theta A}P-e^{\theta PAP}P\|_2
&\le\cosh(|\theta|\lambda)-1.
\end{align}
These are uniform operator bounds: they hold for every normalized input state in the target subspace, but can therefore be conservative for a particular trajectory.

For a sequence of parent generators,
\begin{equation}
\mathcal U_{\rm phys}
=
e^{\theta_MA_M}\cdots e^{\theta_1A_1},
\end{equation}
define $A_{0,k}=PA_kP+QA_kQ$ and the corresponding block-diagonal reference sequence
\begin{equation}
\mathcal U_0
=
e^{\theta_MA_{0,M}}\cdots e^{\theta_1A_{0,1}}.
\end{equation}
For an initial state in $\operatorname{Ran}(P)$, the $P$-sector action of $\mathcal U_0$ is exactly the lifted reduced-space trajectory.
A Duhamel bound applied factor by factor and a telescoping sum give
\begin{equation}
\|\mathcal U_{\rm phys}-\mathcal U_0\|_2
\le
L
\equiv
\sum_{k=1}^{M}|\theta_k|\,\|QA_kP\|_2 .
\label{eq:L}
\end{equation}
Thus $L\ll1$ is sufficient for representation fidelity.
A large $L$ is not a proof of failure because this uniform bound can be loose and ignores the actual states visited by the ansatz.

\subsection{State-specific telescoping certificate}

A tighter certificate can be obtained by evaluating each factor on the specific lifted projected trajectory rather than taking a worst-case operator norm.
Define
\begin{equation}
R_k=e^{\theta_kA_k},
\qquad
R_{0,k}=e^{\theta_kA_{0,k}},
\end{equation}
and let
\begin{equation}
|\phi_{k-1}\rangle
=
R_{0,k-1}\cdots R_{0,1}|\psi_0\rangle
\end{equation}
be the lifted projected state immediately before factor $k$.
The exact local discrepancy introduced by that factor is
\begin{equation}
\delta_k
=
\|(R_k-R_{0,k})|\phi_{k-1}\rangle\|,
\end{equation}
and the accumulated state-specific budget is
\begin{equation}
D_{\rm seq}=\sum_k\delta_k.
\label{eq:Dseq}
\end{equation}

The exact telescoping identity
\begin{equation}
R_M\cdots R_1-R_{0,M}\cdots R_{0,1}
=
\sum_{k=1}^{M}
R_M\cdots R_{k+1}(R_k-R_{0,k})R_{0,k-1}\cdots R_{0,1}
\end{equation}
together with the unitarity of all $R_j$ and $R_{0,j}$ gives
\begin{equation}
\|\Psi_{\rm phys}-\Psi_{\rm proj}\|
\le D_{\rm seq},
\end{equation}
where
$|\Psi_{\rm phys}\rangle=\mathcal U_{\rm phys}|\psi_0\rangle$
and
$|\Psi_{\rm proj}\rangle=\mathcal U_0|\psi_0\rangle$.
For normalized states, the fidelity is
$F(\Psi_{\rm phys},\Psi_{\rm proj})
=|\langle\Psi_{\rm proj}|\Psi_{\rm phys}\rangle|^2$.
When $D_{\rm seq}\le\sqrt2$,
\begin{equation}
F(\Psi_{\rm phys},\Psi_{\rm proj})
\ge
\left(1-\frac{D_{\rm seq}^2}{2}\right)^2.
\label{eq:Fcert}
\end{equation}
Because $|\Psi_{\rm proj}\rangle\in\operatorname{Ran}(P)$, the target-subspace weight of the physical state,
\[
p_P(\Psi_{\rm phys})
=
\langle\Psi_{\rm phys}|P|\Psi_{\rm phys}\rangle
=
1-\|Q|\Psi_{\rm phys}\rangle\|^2,
\]
obeys
\begin{equation}
p_P(\Psi_{\rm phys})
\ge
\max(0,1-D_{\rm seq}^2).
\label{eq:Pcert}
\end{equation}
Unlike $L$, this certificate is state-specific: it measures the discrepancies encountered along the actual projected trajectory rather than the largest possible action on the entire target subspace.

Table~\ref{tab:statecert} summarizes the molecular negative controls.
The reported diagnostic distance is the phase-minimized pure-state distance
$d_{\rm proj}=\sqrt{2-2|\langle\Psi_{\rm proj}|\Psi_{\rm phys}\rangle|}$.
Because $d_{\rm proj}\le\|\Psi_{\rm phys}-\Psi_{\rm proj}\|$ for any fixed relative phase, it can be compared conservatively with the vector-norm certificate $D_{\rm seq}$ but is not the quantity directly bounded in Eq.~\ref{eq:Dseq}.
\begin{table}[htbp]
\centering
\small
\caption{State-specific sequence certificate. A zero lower bound denotes a mathematically valid but non-informative certificate.}
\label{tab:statecert}
\aipalt{Rows compare the Cu, NO, and OH molecular negative controls. Columns report the state-specific sequence budget $D_{\rm seq}$, the projected-versus-physical state distance $d_{\rm proj}$, the certified minimum fidelity, and the observed fidelity. The Cu certificate is non-informative and the observed fidelity is low, whereas the NO and especially OH certificates are nonzero and increasingly tight.}
\begin{tabular}{lrrrr}
\toprule
System & $D_{\rm seq}$ & $d_{\rm proj}$ & Certified $F_{\min}$ & Actual $F$\\
\midrule
Cu CAS(15e,9o) & 4.6026 & 1.2075 & 0 & 0.07343\\
NO CAS(7e,6o) & 0.3988 & 0.1678 & 0.8473 & 0.97205\\
OH CAS(7e,5o) & 0.06666 & 0.02776 & 0.99556 & 0.999229\\
\bottomrule
\end{tabular}
\end{table}
For NO, Eq.~\ref{eq:Pcert} also certifies target-doublet weight above 0.8410, compared with the observed 0.97211.
For OH the certified lower bound is 0.99556, compared with the observed 0.999229.
For the severe Cu sequence the state-specific sum is substantially smaller than the uniform operator-norm budget $L=12.686$, but it is still too large to provide a nontrivial fidelity certificate.

\section{Controlled single-generator numerical check}

\begin{figure}[htbp]
\centering
\includegraphics[width=0.80\linewidth]{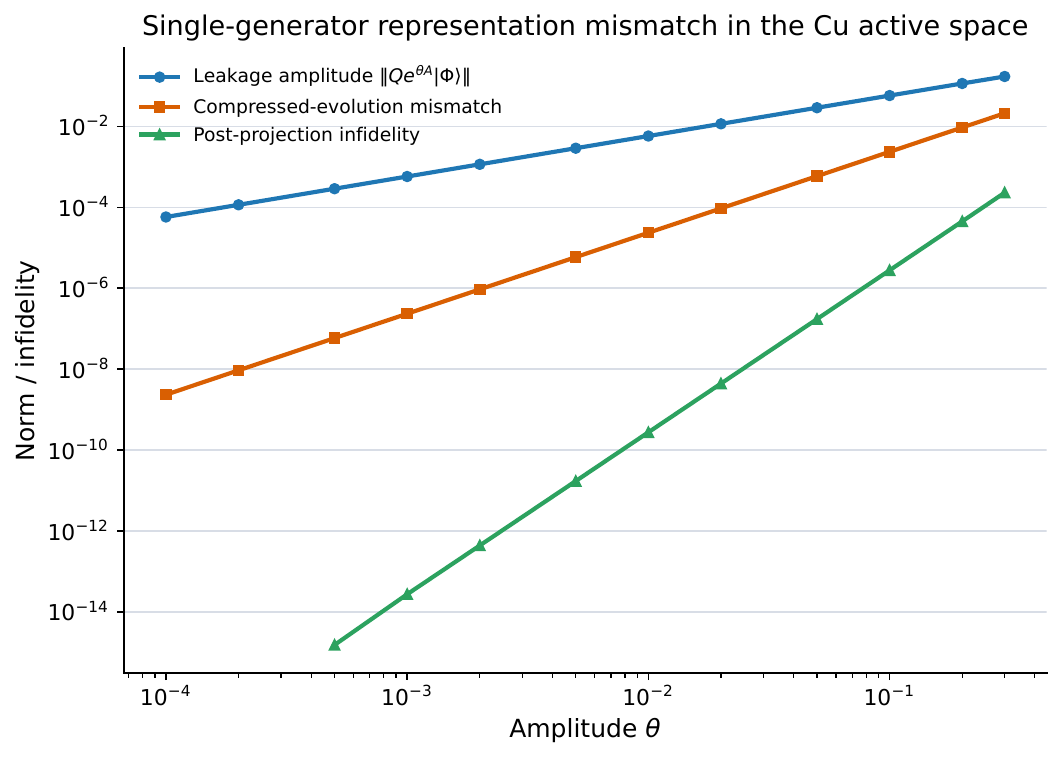}
\caption{Single-generator representation-equivalence test in the Cu active space.}
\label{fig:single_scaling}
\aipalt{Log-log plot of generator amplitude $\theta$ versus three errors for one non-invariant Cu excitation. Leakage amplitude grows approximately linearly with $\theta$, the compressed-evolution mismatch grows quadratically, and the normalized post-projection infidelity falls on an approximately fourth-order trend at small amplitude.}
\end{figure}

For a deterministic bare Cu excitation (remove spin orbital 0, add spin orbital 8),
\begin{equation}
\ell_F=
\frac{\|QAU\|_F}{\|AU\|_F}
=
0.5123475383,
\end{equation}
and
\begin{equation}
\|QAP|\Phi_{\rm HF}\rangle\|
=
0.5773502692.
\end{equation}
Exact determinant-space evolutions over $10^{-4}\le\theta\le0.3$ give small-$\theta$ fitted powers
1.0000 for leakage amplitude,
2.0000 for leakage probability, and
2.0000 for the compressed-evolution mismatch norm.
The normalized post-projection infidelity scales approximately as $\theta^{4.03}$ over the numerically resolved part of the same interval.
The resulting scaling is shown in Fig.~\ref{fig:single_scaling}.

As a preserving control, the first generator of the final physical 18q sequence has $\ell_F=3.57\times10^{-16}$ and remains equivalent to double precision for test amplitudes through $\theta=1$.

\section{Canonical Cu projected-ADAPT negative control}

The main negative control was regenerated from the current canonical CAS(15e,9o) FCIDUMP, not inherited from the superseded historical checkpoint.
The deterministic projected ADAPT run contains 121 operators and reaches
\begin{equation}
E_{\rm proj}=-2518.987495463899~E_h,
\end{equation}
or 1.571906 m$E_h$ above the exact target doublet.
Applying the same selected labels and amplitudes to the bare full-space parent generators gives
\begin{align}
E_{\rm bare}&=-2518.524337694489~E_h,\\
\Delta E_{\rm bare}&=464.729675~{\rm m}E_h,\\
\langle S^2\rangle_{\rm bare}&=2.383536856,\\
F_{\rm proj,bare}&=0.0734304134,\\
p_D^{\rm bare}&=0.455487715 .
\end{align}
In this active space only doublet and quartet spin sectors are available in fixed $M_S=1/2$, so
\begin{equation}
p_Q^{\rm bare}=0.544512285 .
\end{equation}

The normalized post-projected doublet component has
\begin{align}
\Delta E_D^{\rm post}&=465.164029~{\rm m}E_h,\\
F(\Psi_{\rm proj},\Psi_{\rm bare,D})&=0.161212720.
\end{align}
The complementary normalized quartet component lies 470.308070 m$E_h$ above the exact quartet minimum.
Thus the failure is not adequately described as contamination of an otherwise correct target state.

\begin{figure}[htbp]
\centering
\includegraphics[width=0.84\linewidth]{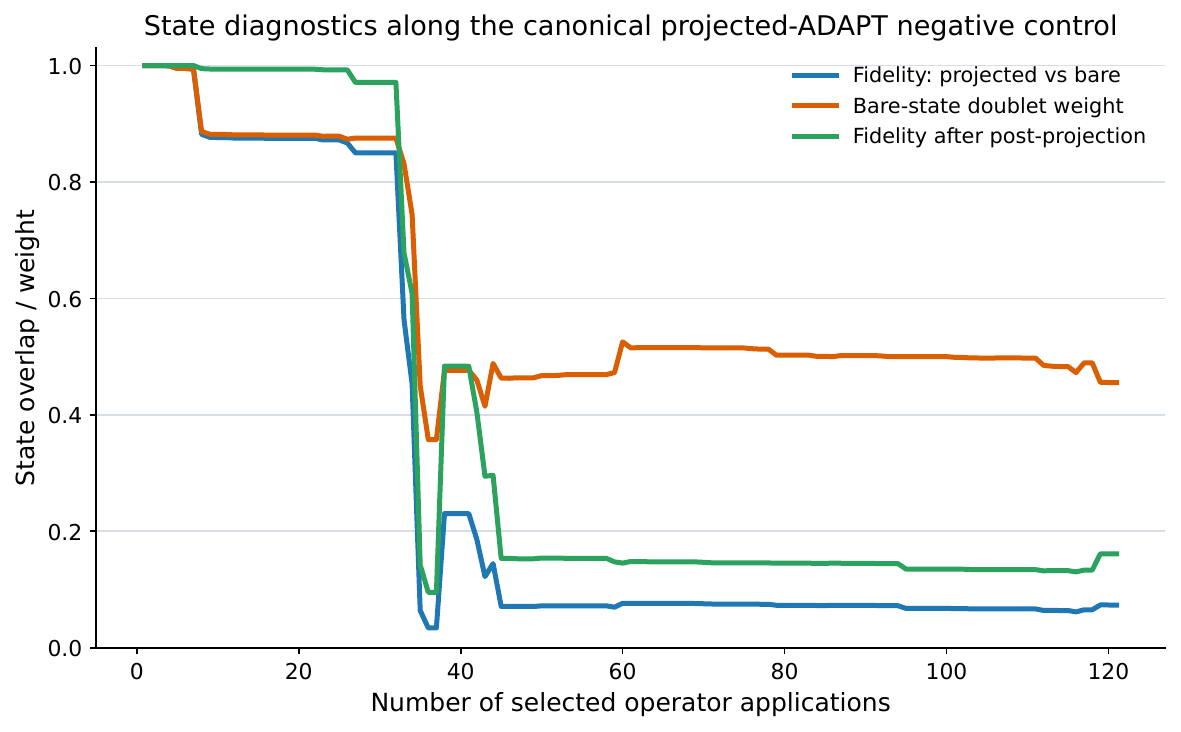}
\caption{Prefix-level state diagnostics for the canonical Cu projected-ADAPT negative control.}
\label{fig:prefix_state}
\aipalt{Line plot over the 121 selected Cu operator applications. The fidelity between the projected and bare full-space states, the bare-state doublet weight, and the fidelity after post-projection all drop sharply near the middle of the sequence and remain far below one for most later prefixes.}
\end{figure}

The 121 selected generators contain 114 non-invariant factors.
The maximum and median normalized Frobenius leakages are 0.557086 and 0.512348, respectively.
The sequence budgets are
\begin{equation}
L=12.68645,\qquad D_{\rm seq}=4.60259.
\end{equation}
The minimum prefix fidelity is 0.03405 and the minimum target-doublet weight is 0.35760.
The corresponding prefix-level state diagnostics are shown in Fig.~\ref{fig:prefix_state}.

\section{Fixed parent-sequence full-space reoptimization}

The severe same-angle Cu mismatch tests representation equivalence, but by itself it does not determine whether the 121 parent generators are intrinsically unable to represent a good target-doublet state after their parameters are reoptimized.
We therefore keep the \emph{same 121 full-space generators in exactly the same order} and optimize all amplitudes directly in the fixed-$M_S=1/2$ determinant space.

The target-spin constraint is imposed on
\begin{equation}
p_D(\boldsymbol\theta)
=
\|U^\dagger|\Psi(\boldsymbol\theta)\rangle\|^2
\end{equation}
using the Sequential Least Squares Programming (SLSQP) algorithm with analytic derivatives of both the physical energy and $p_D$.
All constrained runs use \texttt{ftol}$=10^{-12}$ and at most 1000 SLSQP iterations; the final constraint value is reported explicitly rather than inferred from the optimizer status.
The determinant-excitation exponentials are evaluated exactly as independent two-level rotations in the determinant basis; this representation agrees with \texttt{expm\_multiply} to better than $1.4\times10^{-16}$ in the validation check.
The operator order is never changed.

For the principal constraint $p_D\ge0.999$, four starts were used: the zero-amplitude vector and three independent Gaussian perturbations of it with standard deviation 0.05 (random seed 9272026).
All four optimizations terminate successfully.
Three converge to the same energy within $5\times10^{-10}$ m$E_h$ and the fourth to a nearby local minimum:
The four principal constrained runs are summarized in Table~\ref{tab:fixed_parent_reopt}.
\begin{table}[htbp]
\centering
\small
\caption{Full-space reoptimization of the fixed 121-generator Cu parent sequence under $p_D\ge0.999$. The projected reduced-space result is 1.5719 m$E_h$.}
\label{tab:fixed_parent_reopt}
\aipalt{Four constrained full-space reoptimizations of the same 121-generator Cu parent sequence are compared. Three starts converge to the same 8.481522 m$E_h$ error with $p_D=0.999$ and $F_{\rm proj}=0.979887$; a fourth nearby local minimum is slightly worse at 8.973359 m$E_h$.}
\begin{tabular}{lrrrr}
\toprule
Start & $\Delta E$ (m$E_h$) & $p_D$ & $F_{\rm proj}$ & Post-proj.\ $\Delta E_D$ (m$E_h$)\\
\midrule
Zero & 8.481522 & 0.999000 & 0.979887 & 7.335468\\
Zero + perturbation 1 & 8.481522 & 0.999000 & 0.979887 & 7.335467\\
Zero + perturbation 2 & 8.481522 & 0.999000 & 0.979887 & 7.335471\\
Zero + perturbation 3 & 8.973359 & 0.999000 & 0.977998 & 7.995240\\
\bottomrule
\end{tabular}
\end{table}

The best solution has
\begin{equation}
\langle S^2\rangle=0.753000,\qquad
F(\Psi_{\rm proj},\Psi_{\rm full})=0.979887 .
\end{equation}
Tightening the constraint to $p_D\ge0.9999$ and starting from zero amplitudes yields 11.269613 m$E_h$, with $F_{\rm proj}=0.970168$ and a post-projected error of 11.030494 m$E_h$.

The constrained landscape is nonconvex.
Starts in the basin of the original projected amplitudes converge to substantially poorer stationary solutions (43.7706 m$E_h$ for $p_D\ge0.999$ and 47.8917 m$E_h$ for $p_D\ge0.9999$).
These higher minima are retained as an optimizer diagnostic rather than used to support the main conclusion.

This test changes the interpretation of the catastrophic same-angle result.
The fixed parent sequence is capable of approaching the projected target much more closely after target-spin-constrained full-space reoptimization, so the 464.730 m$E_h$ failure should not be described as evidence that the parent sequence is intrinsically useless.
However, none of the converged constrained searches reaches the 1.6 m$E_h$ benchmark, and the best normalized doublet component remains 7.335 m$E_h$ above the exact target.
The robust conclusion is therefore that the reduced-space optimized amplitudes are not transferable to the parent full-space sequence without an equivalence check; a residual fixed-order expressivity/optimization gap remains in the tested searches, but global optimality is not claimed.

To probe fixed-order representability more directly, we also optimize the same 121 amplitudes for fidelity with the lifted projected state rather than for energy.
The analytic fidelity gradient was independently checked against central finite differences at six amplitudes of the canonical checkpoint; the maximum absolute discrepancy was $2.8\times10^{-11}$.
The objective is
\begin{equation}
\max_{\boldsymbol\theta}
F_{\rm proj}(\boldsymbol\theta)
=
\left|
\langle\Psi_{\rm proj}|\Psi_{\rm full}(\boldsymbol\theta)\rangle
\right|^2 .
\end{equation}
Starting from the best $p_D\ge0.999$ energy-constrained basin and three independent perturbations of it, unconstrained fidelity maximization converges in all four cases to
\begin{equation}
F_{\rm proj}=0.988268496,\qquad p_D=0.997155
\end{equation}
to the displayed precision.
The corresponding energy error is 9.8748 m$E_h$ and the normalized doublet component is 7.7646 m$E_h$ above the exact doublet.

Because that fidelity optimum still contains about $0.28\%$ non-doublet weight, we repeat the optimization with the hard constraint $p_D\ge0.999$.
Three tested starts---the best energy-constrained state, a perturbation of it, and the best unconstrained-fidelity state---all converge to
\begin{equation}
F_{\rm proj}=0.987416189,\qquad p_D=0.999000,
\end{equation}
with energy error 10.8166 m$E_h$.
The agreement across these starts identifies a robust local optimum, not a certified global maximum.
Accordingly, the fidelity test strengthens only the empirical statement that the fixed-order full-space sequence does not reproduce the projected state in the searches performed here; it does not prove that exact reproduction is impossible.

\section{Pool-wide invariance and first-order accessibility}

For each bare Hartree--Fock-origin S/D/T excitation, invariance is tested by $\ell_F=0$ to numerical tolerance.
For the Cu CAS(15e,9o) pool, 274 of 323 bare excitations are non-invariant:
22/22 singles, 126/133 doubles, and 126/168 triples.
Only 49 survive simple invariance filtering.

A useful local expressivity diagnostic is the numerical rank of the vectors obtained by acting on the Hartree--Fock reference and restricting to the exact target space.
The rank tolerance is $10^{-10}$.
Because these Hamiltonians, basis vectors, and anti-Hermitian determinant-excitation matrices are real, this audit concerns the real normalized-state manifold; its maximum nontrivial tangent dimension is therefore $d_D-1$ rather than the complex-projective dimension $2d_D-2$.
This is a local first-order diagnostic only and is not a global controllability theorem.
The resulting tangent-space ranks are summarized in Table~\ref{tab:tangent}.

\begin{table}[htbp]
\centering
\small
\caption{Hartree--Fock target-space tangent accessibility. The maximum nontrivial tangent dimension is $d_D-1$.}
\label{tab:tangent}
\aipalt{Target-space tangent dimensions at the Hartree--Fock reference for Cu, NO, and OH. Projected bare pools span substantially more directions than invariant-only bare pools, while generalized spin-preserving pools recover part of the lost tangent accessibility.}
\begin{tabular}{lrrrr}
\toprule
System & $d_D-1$ & Projected bare S/D/T & Invariant bare & Generalized physical\\
\midrule
Cu CAS(15e,9o) & 239 & 239 & 49 & 99\\
NO CAS(7e,6o) & 209 & 171 & 6 & 56\\
OH CAS(7e,5o) & 39 & 39 & 9 & 25\\
\bottomrule
\end{tabular}
\end{table}

For Cu, simple invariant filtering therefore reduces the first-order accessible target-space rank from the full 239 directions to 49.
The generalized physical rank-1/rank-2 pool restores 99 independent Hartree--Fock tangent directions.
This does not by itself prove global expressivity, but it quantitatively explains why projection can appear highly expressive while simply deleting every leaking bare generator can be too restrictive.

Direct adaptive calculations confirm the point.
Using all 49 invariant Cu bare generators once gives 210.619 m$E_h$ error.
Allowing repeated selection through 100 applications improves only to 167.340 m$E_h$.

\section{Independent NO benchmark}

The NO bare projected S/D/T pool contains 231 generators.
A direct audit finds 225 non-invariant:
17/17 singles, 81/87 doubles, and 127/127 triples.
Projected ADAPT reaches 1.428610 m$E_h$ error in 16 applications.
Eleven of those 16 selected factors are non-invariant.
For the same amplitudes in the bare sequence,
\begin{align}
\Delta E_{\rm bare}&=24.536971~{\rm m}E_h,\\
F_{\rm proj,bare}&=0.972050492,\\
p_D^{\rm bare}&=0.972105986,\\
\langle S^2\rangle_{\rm bare}&=0.833691971.
\end{align}
Here
\begin{equation}
L=0.575942,\qquad D_{\rm seq}=0.398787.
\end{equation}
Unlike the Cu case, post-projection nearly repairs the state:
\begin{align}
\Delta E_D^{\rm post}&=1.477956~{\rm m}E_h,\\
F(\Psi_{\rm proj},\Psi_{\rm bare,D})&=0.999942914.
\end{align}

The lowest exact doublet is twofold degenerate.
The optimized projected state has 0.997999 weight in this manifold, the bare state 0.970221, and the post-projected bare state 0.998061.

Only six of the 231 bare pool elements are invariant.
Using all six once gives 52.154 m$E_h$ error.
Allowing repeats through 60 applications leaves the same energy to the reported precision, showing that simple invariant filtering is again too restrictive.

A generalized physical rank-1/rank-2 pool contains 165 spin-preserving generators.
A 19-application sequence with 13 unique generators reaches
\begin{align}
\Delta E&=1.579026~{\rm m}E_h,\\
\langle S^2\rangle&=0.75,\\
p_D&=1,\\
w_{\rm lowest\ doublet\ manifold}&=0.998967.
\end{align}
The maximum selected representation leakage is below $4.4\times10^{-16}$.

The corresponding 12-qubit Suzuki-2 $r=1$ Qiskit circuit has 7,328 controlled-X (CX) gates, depth 9,869, synthesis-energy error 0.002927 m$E_h$, and circuit-to-fermionic fidelity 0.999993871.
It therefore passes the same strict circuit-equivalence criteria used for the Cu circuits.

\section{Auxiliary OH stress test}

For OH CAS(7e,5o), projected ADAPT reaches 1.05508 m$E_h$ in only five factors.
Three are non-invariant, yet the sequence budget is small:
\begin{equation}
L=0.0670781,\qquad D_{\rm seq}=0.0666610.
\end{equation}
The bare same-angle sequence remains close to the projected result:
\begin{align}
\Delta E_{\rm bare}&=1.93276~{\rm m}E_h,\\
F_{\rm proj,bare}&=0.999229332,\\
p_D^{\rm bare}&=0.999229500.
\end{align}
Post-projection yields 1.04156 m$E_h$ error and fidelity 0.999999832 with the projected state.
This provides a useful counterexample to any interpretation that a nonzero invariance defect must always produce a catastrophic error.

The generalized physical pool contains 80 generators and has machine-precision representation leakage.
However, the present first-gradient ADAPT search stalls after repeated reuse of only six unique generators.
After 100 applications,
\begin{align}
\Delta E&=2.58052~{\rm m}E_h,\\
p_D&=1,\\
w_{\rm lowest\ doublet\ manifold}&=0.997527,\\
r&=0.05546~E_h,
\end{align}
while the maximum available first derivative is only $3.9\times10^{-7}$.
The available archived OH closure summary did not preserve the numerical rank tolerance and implementation details required for a reproducible Lie-algebra claim, so no controllability conclusion is drawn here.
The robust conclusion is narrower: a representation-faithful physical pool can still be difficult for this first-gradient adaptive search.
For this reason OH is not counted as a successful physical-ADAPT benchmark in the main text.

\section{Hubbard-model representation stress tests}

The same projected-versus-bare audit was applied to the homogeneous four-site, five-electron open Hubbard chain.
The projected calculation reaches the exact target doublet to numerical precision for every tested $U/t$.
The physical same-angle sequence does not.
The Hubbard results are summarized in Table~\ref{tab:hubbard}.

\begin{table}[htbp]
\centering
\small
\caption{Hubbard-chain stress tests. Energies are in units of $t$.}
\label{tab:hubbard}
\aipalt{Four-site, five-electron Hubbard-chain representation tests for $U/t=1,2,4,8$. As $U/t$ increases, the bare same-angle energy error increases and the projected-versus-bare fidelity decreases, while the target-doublet weight remains high but imperfect and the uniform leakage budget $L$ is large.}
\begin{tabular}{rrrrrr}
\toprule
$U/t$ & Ops & Bare error & $F_{\rm proj,bare}$ & $p_D$ & $L$\\
\midrule
1 & 19 & 0.2661 & 0.9244 & 0.9574 & 4.858\\
2 & 19 & 0.5462 & 0.8673 & 0.9656 & 9.149\\
4 & 20 & 1.3914 & 0.6333 & 0.8815 & 9.738\\
8 & 20 & 2.8709 & 0.4763 & 0.9359 & 13.091\\
\bottomrule
\end{tabular}
\end{table}

\begin{figure}[htbp]
\centering
\includegraphics[width=0.80\linewidth]{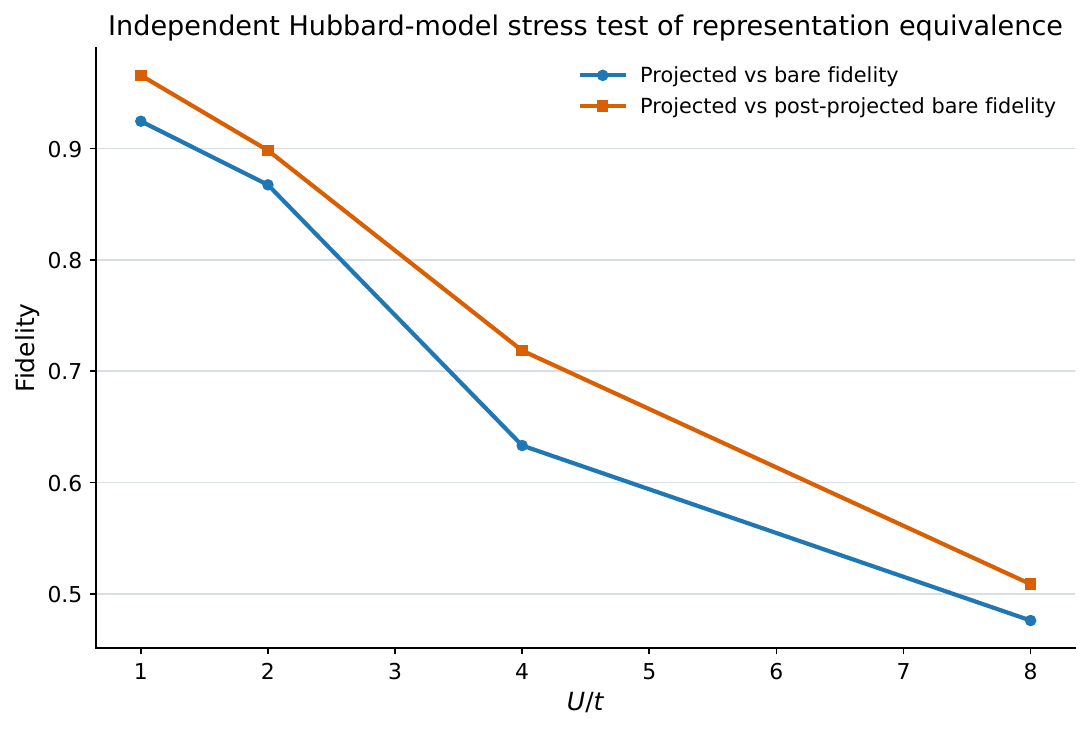}
\caption{Representation-equivalence failure in the homogeneous Hubbard-chain stress tests.}
\label{fig:hubbard}
\aipalt{Line plot of fidelity versus $U/t$ for the Hubbard-chain stress tests. Both the projected-versus-bare fidelity and the projected-versus-post-projected fidelity decrease as interaction strength increases; post-projection improves the fidelity at every tested $U/t$ but does not restore unity.}
\end{figure}

The corresponding fidelity trends are shown in Fig.~\ref{fig:hubbard}.

These small models are not offered as evidence for chemical relevance.
Their purpose is to show that the representation issue does not rely on the detailed Cu or molecular Hamiltonian.

\section{Full-space spin-preserving operator pools}

This construction is used only as a representation-faithful positive control for the audit.
It is not claimed as a new or preferred scalable spin-adapted ADAPT algorithm; symmetry-preserving state preparation, spin-adapted pools, symmetry-completeness analyses, and exact spin-adapted factorizations are established elsewhere \cite{gard2020symmetry,shkolnikov2023roadblocks,burton2023disco,magoulas2026symmetry,magoulas2025closedform,jain2026factorization}.
Its role here is to provide a full-space sequence for which target-sector preservation can be checked independently and the reduced/full-space trajectories can therefore be compared without representation ambiguity.

A determinant excitation is represented by
\begin{equation}
A=T-T^\dagger.
\end{equation}
Generalized rank-1 and rank-2 $M_S$-preserving excitation patterns are grouped by their spatial remove/add structure.
Within each group, linear combinations
\begin{equation}
B=\sum_j c_jA_j
\end{equation}
are obtained from the null space of the coupling from the target spin space to its complement.
Candidates are retained only when both their target-space leakage and normalized $S^2$ commutator defect lie below numerical tolerance.

For Cu 18q, the complete generalized pool contains 1,080 generators (36 rank-1 and 1,044 rank-2).
The maximum full-pool leakage and normalized $S^2$ commutator defect are at the $10^{-15}$ level.
For Cu 24q, the corresponding pool contains 3,762 generators (66 rank-1 and 3,696 rank-2).
For NO the generalized physical pool contains 165 generators, and for OH 80.

\subsection{Full-sequence positive-control audit}

The final full-space spin-preserving Cu sequences were reconstructed independently in the full fixed-$M_S$ space and in the exact-doublet basis.
For 18q:
\begin{align}
\max_k\ell_F(B_k)&=4.13\times10^{-16},\\
\sum_k|\theta_k|\|QB_kU\|_F&=3.91\times10^{-14},\\
\min_{\rm prefixes}F_{\rm full,reduced}
&=0.9999999999999993.
\end{align}
For 24q:
\begin{align}
\max_k\ell_F(B_k)&=5.17\times10^{-16},\\
\sum_k|\theta_k|\|QB_kU\|_F&=5.14\times10^{-14},\\
\min_{\rm prefixes}F_{\rm full,reduced}
&=0.9999999999999987.
\end{align}
The final full/reduced energies agree to the reported double precision in both cases.
The corrected full-space spin-preserving construction therefore passes the same representation audit that the projected bare sequence fails.

\subsection{Normalization-controlled null-space basis sensitivity}

The spin-preserving coefficient space within one spatial remove/add group can have dimension larger than one, so its individual basis vectors are not physically unique.
The production implementation nevertheless needs a deterministic basis for standard single-operator ADAPT selection and therefore constructs a sparse canonical basis by searching small supports first, fixing a deterministic sign, and snapping coefficients only when they are already within $5\times10^{-10}$ of the exact values $0$, $\pm1$, or $\pm1/\sqrt2$.

A direct normalization audit revealed an important subtlety.
For all 378 two-dimensional Cu 18q null-space groups, the two canonical coefficient vectors have unit Euclidean norm but are not mutually orthogonal:
\begin{equation}
|c_1^\mathsf{T}c_2|=\frac13.
\end{equation}
Likewise, the corresponding determinant-space generators have equal Frobenius norms within numerical precision but normalized Frobenius overlap
\begin{equation}
\frac{|\langle B_1,B_2\rangle_F|}{\|B_1\|_F\|B_2\|_F}=0.3913043478.
\end{equation}
Consequently, applying an $O(2)$ matrix directly to the pair $(B_1,B_2)$ does not preserve the norm of each mixed generator.
Across these groups, such raw mixing can change the Frobenius norm by a factor as large as 1.512 and the Hartree--Fock action norm by a factor as large as 1.732.
Because ADAPT gradients scale with generator normalization, an unnormalized mixing is therefore not a clean test of basis sensitivity.

We instead perform a normalization-controlled prefix audit.
For each of the 51 states along the stored 50-application Cu trajectory (including the Hartree--Fock state), we generate 100 independent random mixes of every two-dimensional null-space group,
\begin{align}
\widetilde B_1&=\cos\alpha\,B_1+\sin\alpha\,B_2,\\
\widetilde B_2&=-\sin\alpha\,B_1+\cos\alpha\,B_2,
\end{align}
and renormalize each mixed direction before comparing the largest ADAPT gradient.
Two conventions are tested independently:
(i) unit Euclidean norm of the underlying coefficient vector and
(ii) the canonical determinant-space Frobenius norm.
All one-dimensional groups are unchanged.

For a metric with Gram matrix $M_{ij}=\langle B_i,B_j\rangle_M$ and signed gradient vector $g_i$, the steepest normalized direction available inside one null-space group has the basis-invariant score
\begin{equation}
G_{\mathcal G}=\sqrt{\boldsymbol g^\mathsf{T}M^{-1}\boldsymbol g}.
\label{eq:group_gradient}
\end{equation}
Under an invertible basis change $B'=BR$, one has $M'=R^\mathsf{T}MR$ and $\boldsymbol g'=R^\mathsf{T}\boldsymbol g$, so Eq.~\ref{eq:group_gradient} is invariant.
The normalization-controlled basis audit is summarized in Table~\ref{tab:basis_sensitivity}.

\begin{table}[htbp]
\centering
\small
\caption{Normalization-controlled basis sensitivity evaluated on all 51 stored Cu 18q prefix states, with 100 independent random bases per prefix. Sensitive prefixes are prefixes for which at least one random basis changes the highest-gradient null-space group relative to the canonical pool.}
\label{tab:basis_sensitivity}
\aipalt{Normalization-controlled basis-sensitivity audit over 51 Cu prefix states. Only 4 of 51 prefixes are sensitive under either coefficient-Euclidean or generator-Frobenius normalization, and the canonical highest-gradient group agrees with the basis-invariant group optimum at 50 of 51 and 49 of 51 prefixes, respectively.}
\begin{tabular}{lccc}
\toprule
Metric & Sensitive & Max.\ changed & Group-opt.\ agreement\\
& prefixes & fraction & (prefixes)\\
\midrule
Coefficient Euclidean & 4/51 & 0.14 & 50/51\\
Generator Frobenius & 4/51 & 0.12 & 49/51\\
\bottomrule
\end{tabular}
\end{table}

The controlled result is therefore more modest than a raw basis mixing would imply.
Standard single-operator ADAPT is formally basis dependent, and a few late or near-degenerate decisions can change under normalized rotations, but the highest-gradient group is robust at most sampled prefix states of the stored Cu trajectory.
At the most sensitive coefficient-normalized prefix, 14 of 100 random bases change the top group; under Frobenius normalization the maximum corresponding fraction is 12 of 100.
The canonical single-operator choice agrees with the coefficient-metric invariant group optimum at 50 of 51 prefixes and with the Frobenius-metric optimum at 49 of 51.

These tests do not establish basis-independent full-trajectory convergence because a single altered choice can change all subsequent optimized states.
They do, however, show that the canonical sparse basis does not produce a large normalization-induced selection bias along the stored trajectory.
The reported 50-application passing state remains a result for the documented canonical basis and search/pruning history, not a proof of a basis-independent minimal ansatz.
A group-wise ADAPT rule based on Eq.~\ref{eq:group_gradient} would remove the arbitrary basis choice once a generator normalization metric is specified.

\section{ADAPT selection, repeats, reoptimization, and method comparisons}

For anti-Hermitian candidate $B_k$,
\begin{equation}
g_k
=
\left.\frac{\partial E}{\partial\theta_k}\right|_0
=
\langle\Psi|[\hat H,B_k]|\Psi\rangle
=
2\,{\rm Re}\langle\Psi|\hat H B_k|\Psi\rangle.
\end{equation}
The largest $|g_k|$ is selected, and the pool is not drained, so previously used generators may be selected again, following the original ADAPT construction \cite{grimsley2019adapt}.
Parameters are optimized with L-BFGS-B, including global reoptimization.
Small-amplitude applications are subsequently pruned and the remaining amplitudes reoptimized. This is used only as a practical compaction step and is not presented as a new pruning algorithm \cite{vaquero2025pruned}.

The final Cu 18q sequence has 50 applications and 39 unique generators.
The 24q sequence has 61 applications and 41 unique generators.
The sequence lengths are the smallest passing states found along the present search/pruning histories and are not proofs of global minimality.
The physical-pool comparison is summarized in Table~\ref{tab:poolcomparison}.

\begin{table}[htbp]
\centering
\small
\caption{Cu 18q physical-pool comparison.}
\label{tab:poolcomparison}
\aipalt{Comparison of four Cu 18-qubit spin-pure operator-pool strategies. Restricted S+D and S+D+T pools have large energy errors; allowing repeats improves the restricted S+D+T result to 13.166 m$E_h$, while the generalized rank-1/rank-2 pool with repeats reaches the lowest error, 1.493 m$E_h$, with 50 applications and fidelity 0.9826.}
\begin{tabular}{lrrrr}
\toprule
Pool & Applications & Error (m$E_h$) & Fidelity & Residual ($E_h$)\\
\midrule
Restricted S+D & 35 & 153.629 & 0.2866 & 0.4189\\
Restricted S+D+T & 147 & 141.392 & 0.2672 & 0.3118\\
Restricted S+D+T + repeats & 250 & 13.166 & 0.8783 & 0.0412\\
Generalized rank-1/rank-2 + repeats & 50 & 1.493 & 0.9826 & 0.0480\\
\bottomrule
\end{tabular}
\end{table}

All rows are spin pure to numerical precision.
This comparison separates spin correctness from practical expressivity.

\section{Qubit mapping and Suzuki synthesis}

Each physical fermionic generator $B_k$ is Jordan--Wigner mapped.
Writing
\begin{equation}
H_k=-iB_k
\end{equation}
gives
\begin{equation}
e^{\theta_kB_k}=e^{-i(-\theta_k)H_k}.
\end{equation}
For
\begin{equation}
H=\sum_{j=1}^{m}H_j,
\end{equation}
the second-order Suzuki step is
\begin{equation}
S_2(t)=
\left(\prod_{j=1}^{m-1}e^{-iH_jt/2}\right)
e^{-iH_mt}
\left(\prod_{j=m-1}^{1}e^{-iH_jt/2}\right).
\end{equation}
With $r$ repetitions, $S_2(t/r)^r$ is used.

The stored validation environment uses Qiskit 2.2.3, Qiskit Aer 0.17.2, and Qiskit Nature 0.7.2.
Circuits are transpiled at optimization level 1 to the abstract basis
$\{R_z,\sqrt X,X,\mathrm{CX}\}$ without a device coupling map.
The original uniform-repetition campaign did not pin \texttt{seed\_transpiler}; those CX/depth values should therefore be read as archived Qiskit-2.2.3 realizations rather than seed-independent compiler invariants.
The selective-refinement confirmation reported below was subsequently re-transpiled with the same Qiskit version and basis using the fixed value \texttt{seed\_transpiler}=9272026.

The strict circuit-equivalence gate is defined after fixing the synthesis-energy tolerance at
0.1 m$E_h$, i.e. 6.25\% of the 1.6 m$E_h$ model-space benchmark, so that product-formula error
remains a subdominant contribution to the target energy scale. This threshold was fixed
before the selective-refinement schedule search.
\begin{align}
F_{\rm circuit,fermionic}&\ge0.999,\\
|\delta E_{\rm synth}|&\le0.1~{\rm m}E_h,\\
|\langle S^2\rangle-0.75|&\le0.001,\\
p(M_S{\rm\ target})&\ge0.999.
\end{align}

\subsection{Local factorization-error audit}

A finite Suzuki product can violate spin even when the complete fermionic generator commutes with $S^2$, because the constituent Pauli or determinant-excitation terms need not commute with one another or with $S^2$ separately.
We therefore evaluate each ansatz application in isolation at the actual optimized amplitude.

For one Suzuki-2 repetition, the maximum isolated-application infidelity is
\begin{align}
0.1610 &\quad\text{for Cu 18q},\\
2.46\times10^{-4} &\quad\text{for Cu 24q},\\
4.19\times10^{-6} &\quad\text{for NO}.
\end{align}
The 18q sum is dominated by one rank-2 application with $|\theta|\simeq3.13$.

Write one serialized physical generator as $B=\sum_a c_aA_a$, where $A_a$ are its nonzero determinant-excitation constituents (terms with $|c_a|\le10^{-12}$ are ignored for this diagnostic).
In the fixed-$N$, fixed-$M_S$ determinant representation we define
\begin{equation}
c_{\rm comm}(B)
=
\max_{a<b}
\frac{\|[A_a,A_b]\|_F}
{\|A_a\|_F\,\|A_b\|_F},
\label{eq:ccomm}
\end{equation}
with $c_{\rm comm}=0$ for a single active constituent.
This quantity is deliberately simple: it is computed before Jordan--Wigner mapping, ignores the constituent coefficients $c_a$, and is therefore not a rigorous error estimator for the Pauli-level Suzuki formula used by Qiskit.

\begin{figure}[htbp]
\centering
\includegraphics[width=0.80\linewidth]{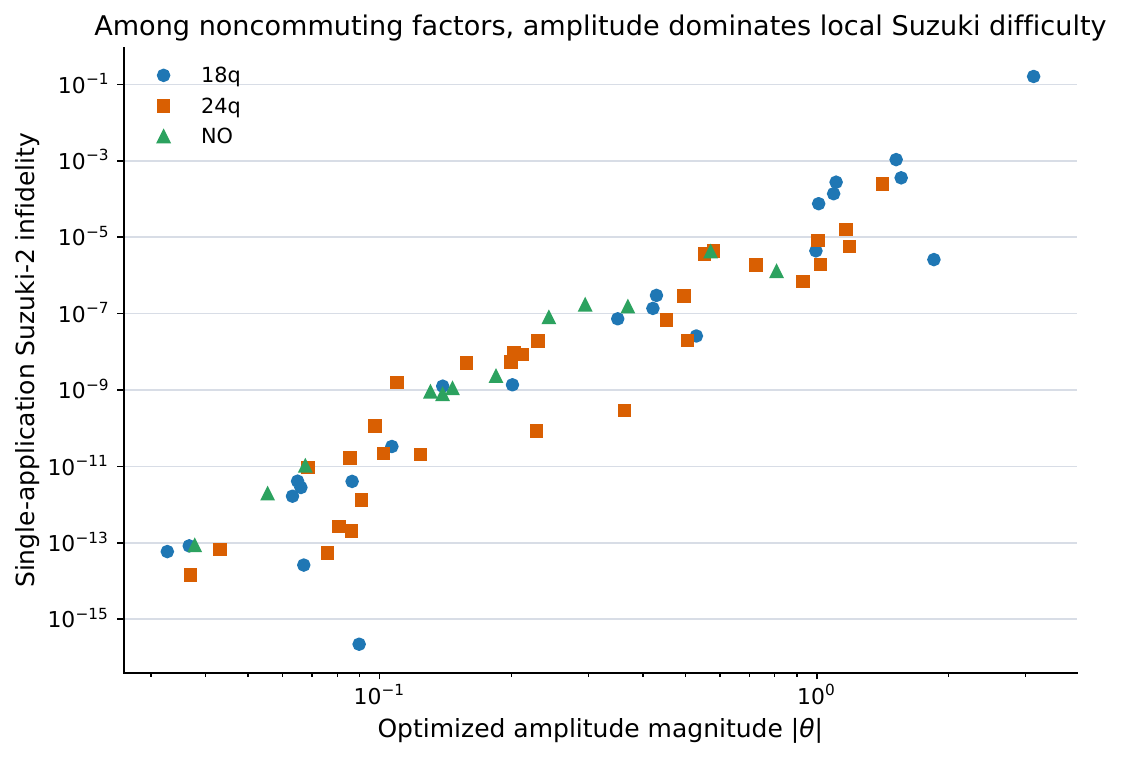}
\caption{Isolated Suzuki-2 infidelity versus optimized amplitude magnitude for applications with $c_{\rm comm}>0$. The amplitude is a strong rank predictor within this noncommuting subset; the continuous $c_{\rm comm}$ factor provides only a modest additional change in the rank correlations reported in the text.}
\label{fig:suzuki_local}
\aipalt{Log-log scatter plot of optimized amplitude magnitude versus isolated Suzuki-2 infidelity for noncommuting factors in the Cu 18-qubit, Cu 24-qubit, and NO ansatzes. Infidelity generally increases strongly with amplitude magnitude, with the largest 18-qubit amplitude also producing the largest isolated error.}
\end{figure}

A direct baseline comparison changes the interpretation of this diagnostic.
For Cu 18q, Cu 24q, and NO, respectively, 24/50, 30/61, and 5/19 applications have $c_{\rm comm}=0$; the largest isolated Suzuki infidelity in those commuting subsets is only $4.4\times10^{-16}$, $1.6\times10^{-15}$, and $2.2\times10^{-16}$.
Thus constituent noncommutativity is an excellent structural separator in these data.
However, inside the subset with $c_{\rm comm}>0$, the Spearman correlations of $|\theta|^3$ alone with isolated-application infidelity are already 0.925, 0.958, and 0.953.
Multiplying by the continuous $c_{\rm comm}$ changes those correlations only to 0.928, 0.970, and 0.953.
We therefore do \emph{not} claim that the numerical magnitude of $c_{\rm comm}$ adds an independent quantitative error law.
The supported conclusion is narrower: commuting constituent sets are easy at this level, while among noncommuting factors the optimized amplitude magnitude carries most of the observed ranking information.
The corresponding noncommuting-factor trend is shown in Fig.~\ref{fig:suzuki_local}.

\subsection{Selective-refinement search and Qiskit confirmation}

Because the 18q error is concentrated in a few factors, a determinant-equivalent Suzuki emulator was first used to search a restricted nonuniform repetition grid.
The six variables were applications 9, 19, 21, 25, 33, and 34, chosen from the largest isolated-application errors; every other application was fixed at $r=1$.
The tested repetition sets were
$r_9\in\{1,2,4,8,16\}$,
$r_{25}\in\{1,2,4,8\}$, and
$r_{19},r_{21},r_{33},r_{34}\in\{1,2,4\}$,
for an exhaustive grid of 1620 schedules.
Of these, 277 satisfy the strict state-level criteria.
The lowest component-exponential proxy among this explicitly tested grid refines only applications 9, 21, 25, and 34:
\begin{equation}
r_9=4,\qquad r_{21}=r_{25}=r_{34}=2,
\end{equation}
with all other applications at $r=1$.
The determinant-equivalent emulator gives
\begin{align}
|\delta E_{\rm synth}|&=0.09502~{\rm m}E_h,\\
F_{\rm circ,fermionic}&=0.9995825,\\
\langle S^2\rangle&=0.750724,\\
p_D&=0.999759.
\end{align}
Its component-exponential proxy is 1600, compared with 1334 for uniform $r=1$ and 5336 for uniform $r=4$.

We then re-transpiled \emph{exactly this nonuniform schedule} with Qiskit 2.2.3, Qiskit Aer 0.17.2, and Qiskit Nature 0.7.2, using optimization level 1, the abstract basis
$\{R_z,\sqrt X,X,\mathrm{CX}\}$, and \texttt{seed\_transpiler}=9272026.
The actual transpiled state gives
\begin{align}
|\delta E_{\rm synth}|&=0.0950207827~{\rm m}E_h,\\
F_{\rm circ,fermionic}&=0.9995824567,\\
\langle S^2\rangle&=0.7507241840,\\
p(M_S\text{ target})&=0.9999999999994,\\
p_D&=0.9997586053.
\end{align}
Thus all four predeclared circuit-equivalence criteria are satisfied.
The corresponding abstract-gate-set circuit contains 19,660 CX gates at depth 23,016.
For comparison, uniform $r=4$ contains 61,482 CX gates at depth 71,636 while giving
$|\delta E_{\rm synth}|=0.135462$ m$E_h$ and therefore failing the strict energy gate.
The selective compilation reduces the reported CX count by 68.0\% and depth by 67.9\% relative to that uniform-$r=4$ realization while also passing the state-level gate.

The close agreement with the determinant-equivalent emulator validates the restricted schedule search for this stored ansatz; it does not establish that the schedule is globally resource-optimal.
The counts remain abstract-basis compilation results without a device coupling map and are not hardware-resource claims.
The exact input hashes, fixed compiler seed, and revision-specific validation record are archived with the repository.

Recent exact and closed-form spin-adapted factorizations \cite{magoulas2025closedform,jain2026factorization} remain important alternatives.
We have not established a one-to-one compiled comparison for the particular numerical linear-combination generators used here.

\section{Measurement grouping and shot allocation}

The Cu 18q and 24q Jordan--Wigner Hamiltonians contain 9,316 and 29,737 Pauli terms.
Greedy qubit-wise commuting (QWC) grouping gives 1,909 and 6,233 groups.
For
\begin{equation}
\hat H=\sum_g\hat H_g,\qquad
V_g={\rm Var}_\Psi(\hat H_g),
\end{equation}
and independent group sampling,
\begin{equation}
\sigma_E^2=\sum_g\frac{V_g}{n_g}.
\end{equation}
Variance-optimal allocation gives
\begin{equation}
n_g\propto\sqrt{V_g},
\qquad
N_{\rm shot}^{\rm opt}
=
\frac{\left(\sum_g\sqrt{V_g}\right)^2}{\sigma_E^2}.
\end{equation}
This is optimal only for the committed grouping and ideal state preparation \cite{verteletskyi2020measurement}.
The resulting shot estimates are listed in Table~\ref{tab:shots}.

\begin{figure}[htbp]
\centering
\includegraphics[width=0.80\linewidth]{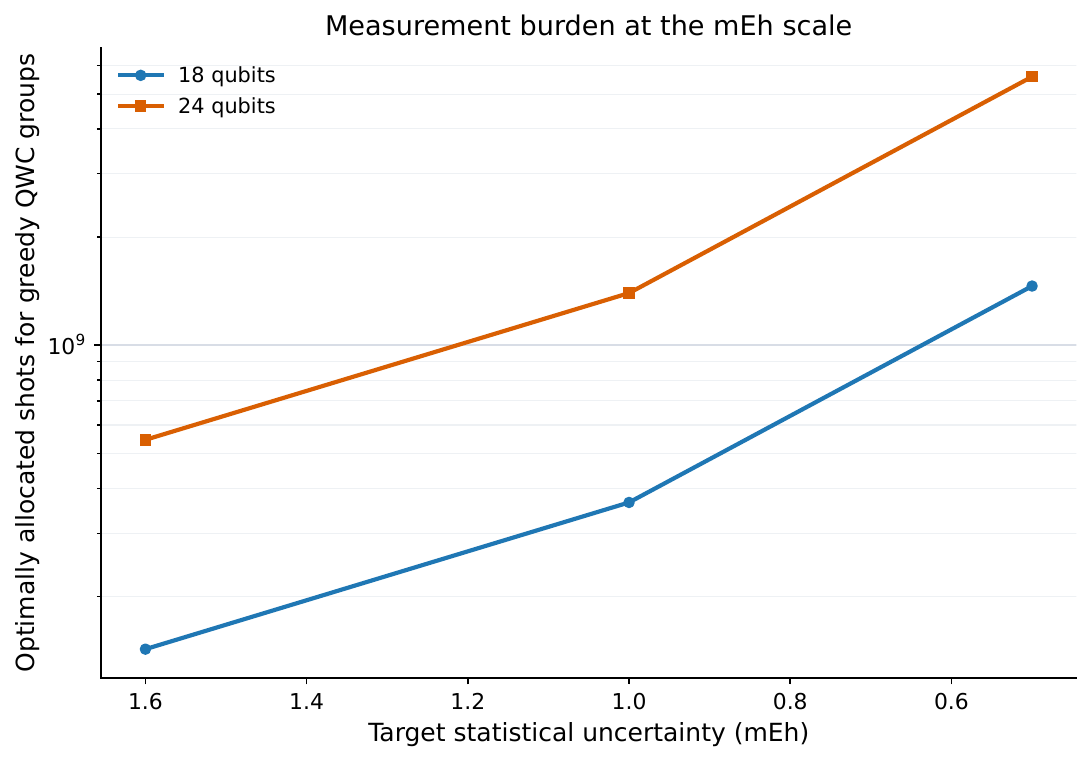}
\caption{Idealized final-energy shot estimates for the chosen greedy QWC groupings. The calculation excludes state-preparation error, device noise, calibration overhead, adaptive-gradient measurements, repeated optimizer evaluations, and alternative measurement strategies.}
\label{fig:si_measurement}
\aipalt{Semi-log line plot of target statistical uncertainty versus optimally allocated final-energy shots. Shot counts rise steeply as the requested uncertainty decreases from 1.6 to 0.5 m$E_h$, with the 24-qubit curve consistently above the 18-qubit curve.}
\end{figure}

\begin{table}[htbp]
\centering
\small
\caption{Idealized final-energy shot estimates for the chosen greedy QWC grouping.}
\label{tab:shots}
\aipalt{Idealized variance-optimally allocated final-energy shot counts for 18- and 24-qubit Cu Hamiltonians at target statistical uncertainties of 1.6, 1.0, and 0.5 m$E_h$. Tighter uncertainty requires rapidly increasing shot counts, and the 24-qubit case is more expensive at every target.}
\begin{tabular}{lrrr}
\toprule
& 1.6 m$E_h$ & 1.0 m$E_h$ & 0.5 m$E_h$\\
\midrule
18 qubits & $1.43\times10^8$ & $3.65\times10^8$ & $1.46\times10^9$\\
24 qubits & $5.46\times10^8$ & $1.40\times10^9$ & $5.59\times10^9$\\
\bottomrule
\end{tabular}
\end{table}
At 1.6 m$E_h$, variance-optimal allocation reduces the count relative to equal allocation by factors 26.66 and 42.91 for 18q and 24q, respectively.
These estimates cover final-energy measurement only, not ADAPT gradient rounds or repeated optimizer evaluations \cite{ramoa2025resources}.
Their dependence on target statistical uncertainty is shown in Fig.~\ref{fig:si_measurement}.

\section{One-particle diagnostics}

\begin{figure}[htbp]
\centering
\includegraphics[width=0.80\linewidth]{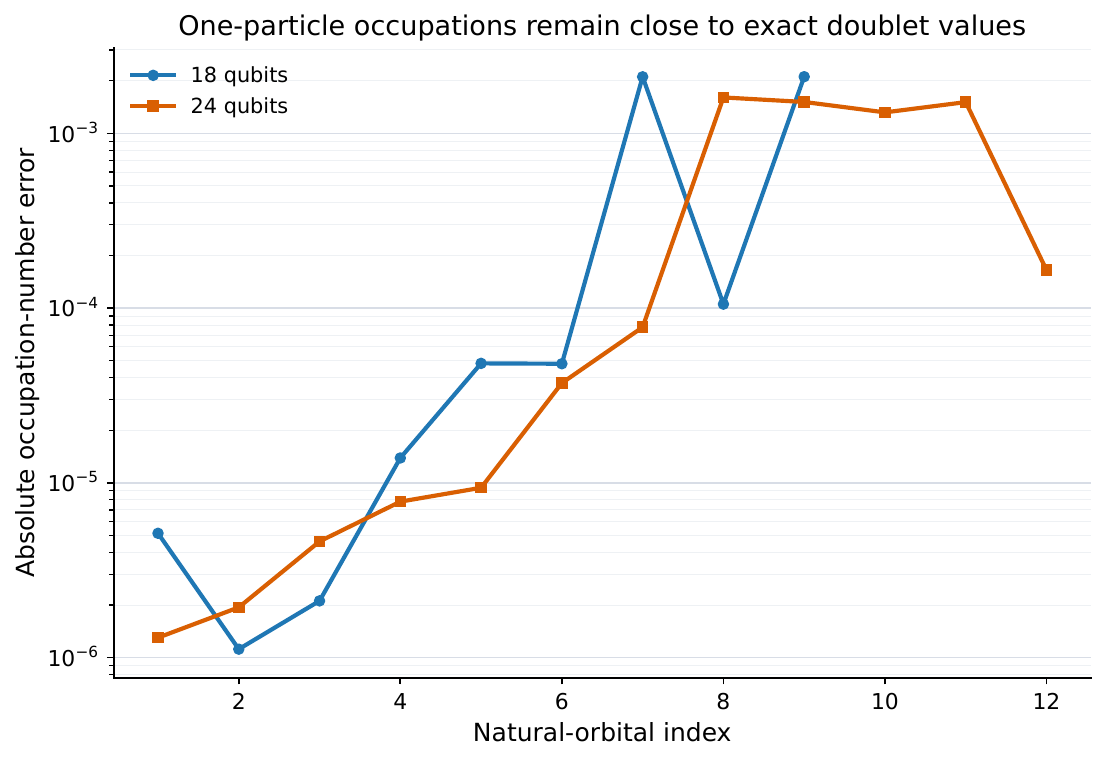}
\caption{Absolute natural-orbital occupation errors of the optimized Cu physical states relative to the exact target doublets.}
\label{fig:natorb}
\aipalt{Log-scale plot of absolute natural-orbital occupation-number errors versus orbital index for the 18- and 24-qubit Cu states. Errors range from roughly one part per million to a few parts in one thousand, with the largest deviations occurring among the higher-index active orbitals.}
\end{figure}

The spin-orbital one-particle reduced density matrix (1-RDM) is
\begin{equation}
\gamma_{pq}
=
\langle\Psi|a_p^\dagger a_q|\Psi\rangle.
\end{equation}
For Cu 18q, the spatial 1-RDM Frobenius error is 0.1860, the maximum natural-occupation error 0.002109, and the occupation-vector $L_2$ error 0.002983.
For 24q, the corresponding values are 0.1250, 0.001605, and 0.002985.
The orbital-resolved occupation-number errors are shown in Fig.~\ref{fig:natorb}.

The full-state residuals remain 0.0480 $E_h$ and 0.0585 $E_h$, so the compact states should not be interpreted as exact wavefunctions despite their model-space energy accuracy.

\section{Multi-start robustness}

\begin{figure}[htbp]
\centering
\includegraphics[width=0.80\linewidth]{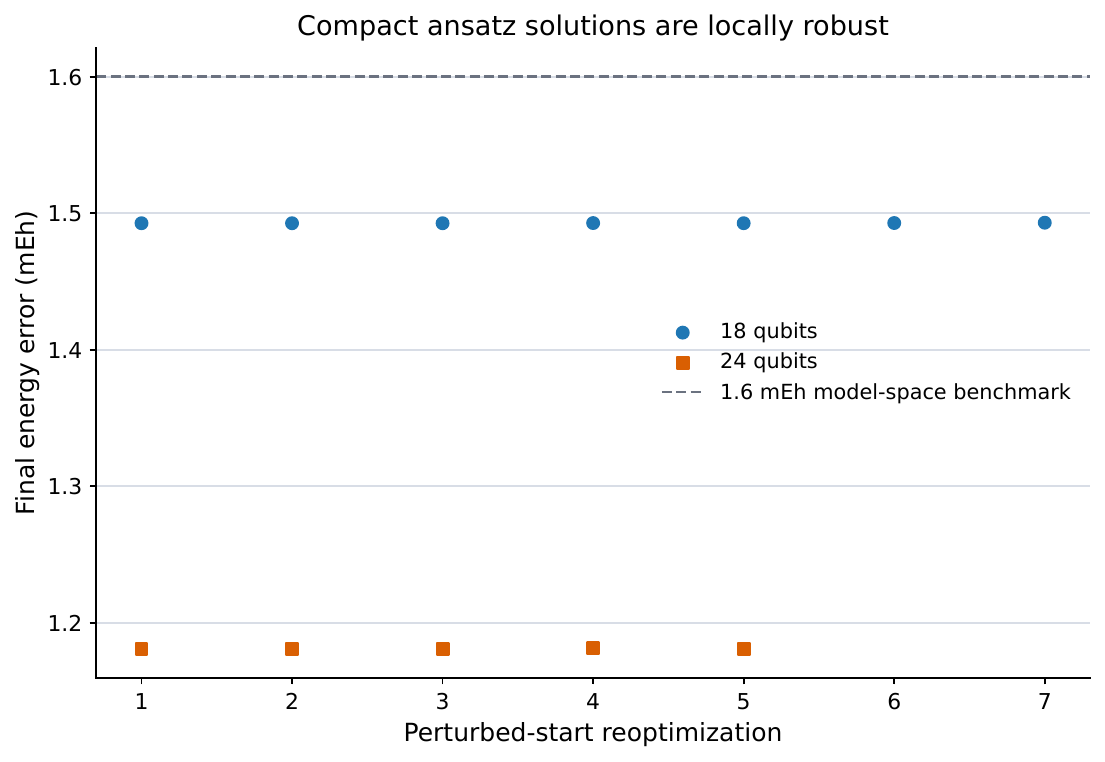}
\caption{Energy errors after reoptimization from perturbed parameter vectors.}
\label{fig:multistart}
\aipalt{Scatter plot of final energy error for perturbed-start reoptimizations of the compact Cu ansatzes. The 18-qubit results cluster tightly near 1.493 m$E_h$ and the 24-qubit results near 1.181 m$E_h$, both below the dashed 1.6 m$E_h$ benchmark line.}
\end{figure}

Perturbed-start global reoptimizations finish in a narrow 1.49266--1.49306 m$E_h$ interval for Cu 18q and 1.18070--1.18125 m$E_h$ for Cu 24q.
Several L-BFGS-B runs reach their iteration cap.
These tests therefore support local energetic robustness but do not prove global optimizer convergence or uniqueness.
The individual perturbed-start outcomes are shown in Fig.~\ref{fig:multistart}.

\FloatBarrier

\section{Implementation regression tests}

Five-operator circuit prefixes were used to check fermionic signs, Jordan--Wigner conventions, and circuit assembly.
The Cu 18q and 24q prefixes have circuit-to-fermionic fidelities above 0.999999998 at one Suzuki repetition.

As a negative resource control, the restricted 147-operator Cu S+D+T sequence can itself be synthesized accurately at one Suzuki repetition, but requires 218,502 CX gates at depth 261,514 while its fermionic error remains 141.392 m$E_h$.
A circuit may therefore be a faithful implementation of a poor ansatz.

\section{Excluded Cu--S geometry extension}

The exploratory five-point Cu--S scan failed its pre-specified CASSCF-convergence and active-orbital-continuity gates.
Neither the doublet nor state-specific quartet calculations converged consistently across the full scan, and active-orbital continuity fell below threshold away from the reference geometry.
The scan is excluded from all geometry-dependent scientific claims.

\section{Reproducibility and provenance}

The canonical Cu negative control was freshly regenerated from the current 18q FCIDUMP rather than relying on the missing historical 123-operator checkpoint.
The relevant hashes are listed explicitly below so that the exact files can be identified without relying on filenames alone:

\noindent Cu 18q FCIDUMP:\\
{\small\texttt{4fe2f74860c4a7b1af7e677f62f4d59a1571c2758ff427ba8627607dcd8c8536}}

\noindent 121-operator projected-ADAPT checkpoint:\\
{\small\texttt{d372476940ccb67fd50af104958e36e68f8dddbac40933b9e5b5910e23bd9cd6}}

\noindent 121-operator representation audit:\\
{\small\texttt{f28ee7e78fa790a9c5224bea44b6a5156b1662941be402a1fec712d4facabd92}}

The independent benchmark FCIDUMP hashes used in this revision are:

\noindent NO CAS(7e,6o):\\
{\small\texttt{6c04286a3895f61fef94109baacde752b8ab0c74df0c6642a50919d9f423465b}}

\noindent OH CAS(7e,5o):\\
{\small\texttt{bc9c14bce3d63cda90273cd4c2e3a0ad8138a0be809c635fcb182f7117e5f1e1}}

For the 18q representation-faithful positive control, the archived full 50-application checkpoint has SHA256
\begin{center}
{\small\texttt{594214e506c7e2411e1b3e35d9419ad52bf95cc8d60a0ca4e08292f0ba04e4db}.}
\end{center}
The fixed-seed selective-Qiskit validation record is retained under
\texttt{results/revision\_v12/selective\_suzuki\_qiskit/}.
The transpiled circuit artifact returned by the independent rerun, stored in Qiskit's QPY serialization format, has SHA256
\begin{center}
{\small\texttt{ab917b9c9b91413a366707b7d33043655e50591166c8d47b90da222fd8a61c76}.}
\end{center}

The public repository is
\url{https://github.com/lucia-malickova/New-VQE-validation}.
The current repository state includes the regenerated 121-operator checkpoint and representation audit, the canonical 18q FCIDUMP and both compact and fully serialized best-50 positive-control checkpoints, the NO and OH benchmark FCIDUMPs and metadata, the state-specific certificate and tangent-accessibility scripts and outputs, the normalization-controlled null-space-basis audit, the Hubbard stress-test source data, the exhaustive selective-Suzuki emulator search record, the fixed-seed Qiskit confirmation, the Suzuki diagnostic comparison, and the source data used for the revised figures.
The archived legacy geometry and Hamiltonian-generation scripts are retained for the canonical 18q model; the missing dedicated 24q active-space selection driver is explicitly documented above rather than reconstructed from inference.

\end{document}